\documentclass[preprint,showpacs,preprintnumbers,amsmath,amssymb,superscriptaddress]{revtex4}

\usepackage{graphicx}
\usepackage{dcolumn}
\usepackage{bm}
\usepackage{color}
\usepackage{longtable}
\newcommand{\lb}{\left(}
\newcommand{\rb}{\right)}
\newcommand{\ls}{\left[}
\newcommand{\rs}{\right]}

\newcommand{\br}{\mbox{\boldmath$r$}}
\newcommand{\hatr}{\hat{\br}}

\newcommand{\alp}{\alpha}

\newcommand{\del}{\delta}
\newcommand{\eps}{\epsilon}
\newcommand{\vep}{\varepsilon}

\newcommand{\kap}{\kappa}
\newcommand{\lam}{\lambda}

\newcommand{\sig}{\sigma}

\newcommand{\vph}{\varphi}

\newcommand{\Gam}{\Gamma}
\newcommand{\Del}{\Delta}

\newcommand{\half}{\frac{1}{2}}

\newcommand{\benu}{\begin{enumerate}}
\newcommand{\eenu}{\end{enumerate}}
\newcommand{\beq}{\begin{equation}}
\newcommand{\eeq}{\end{equation}}
\newcommand{\beqn}{\begin{eqnarray}}
\newcommand{\eeqn}{\end{eqnarray}}

\newcommand{\beqd}{\begin{eqnarray*}}
\newcommand{\eeqd}{\end{eqnarray*}}
\newcommand{\bea}{\begin{array}}
\newcommand{\eea}{\end{array}}
\newcommand{\bcen}{\begin{center}}
\newcommand{\ecen}{\end{center}}
\newcommand{\btab}{\begin{tabular}}
\newcommand{\etab}{\end{tabular}}
\newcommand{\bsub}{\begin{subequations}}
\newcommand{\esub}{\end{subequations}}

\newcommand{\bra}{\langle}
\newcommand{\ket}{\rangle}

\newcommand{\beit}{\begin{itemize}}
\newcommand{\enit}{\end{itemize}}

\def\m@thcombine#1#2{%
  \setbox0=\hbox{$#1$}
  \setbox1=\hbox{$#2$}
  \ifdim\wd0>\wd1
    \setbox0=\hbox to\wd1{\hss\box0\hss}
  \else
    \setbox1=\hbox to\wd0{\hss\box1\hss}
  \fi
  \mathop{\vcenter{
    \offinterlineskip\box0\box1}}}
\def\lesim{\m@thcombine<\sim}
\def\gesim{\m@thcombine>\sim}

\begin{document}

\title{
{Pair correlation of giant halo nuclei in continuum Skyrme-Hartree-Fock-Bogoliubov theory}}

\author{Y. Zhang} \thanks{e-mail: yzhangjcnp@pku.edu.cn}
\affiliation{State Key Laboratory of Nuclear Physics and Technology, School of Physics, Peking University, Beijing 100871, China}
\affiliation{Graduate School of Science and Technology, Niigata University, Niigata 950-2181, Japan}
\author{M. Matsuo}
\affiliation{Graduate School of Science and Technology, Niigata University, Niigata 950-2181, Japan}
\affiliation{Department of Physics, Faculty of Science, Niigata University, Niigata 950-2181, Japan}
\author{J. Meng} 
\affiliation{State Key Laboratory of Nuclear Physics and Technology, School of Physics, Peking University, Beijing 100871, China}
\affiliation{School of Physics and Nuclear Energy Engineering, Beihang University, Beijing 100191, China}
\affiliation{Department of Physics, University of Stellenbosch, Stellenbosch, South Africa}

\begin{abstract}

The giant halos predicted in neutron-rich Zr isotopes
with $A=124-138$
are investigated by using the self-consistent
continuum Skyrme Hartree-Fock-Bogoliubov approach,
in which the asymptotic behavior of continuum quasiparticle states
is properly treated by the Green's function method.
We study in detail the neutron pair correlation involved in the giant halo
by analyzing the asymptotic exponential tail of the neutron pair
condensate (pair density) in addition to that of the neutron particle density.
The neutron quasiparticle spectra associated with these giant halo nuclei
are examined.
It is found that  the asymptotic exponential tail of the neutron pair condensate is
dominated by non-resonant continuum quasiparticle states corresponding to
the scattering states with low asymptotic kinetic energy.
This is in contrast to
the asymptotic tail of the neutron density, whose main contributions arise
from the resonant quasiparticle states corresponding to the weakly-bound
single-particle orbits and resonance orbits in the Hartree-Fock potential.

\end{abstract}


\maketitle

\section{\label{sec:intr}Introduction}

The pairing properties in weakly bound nuclei
have drawn a lot of attention since
the first halo phenomenon was discovered in $^{11}$Li~\cite{Tanihata}.
However, the halos observed so far
have only one or two nucleons in the light exotic nuclei.
In order to study the influence of correlations
and many-body effects, it would be very interesting to
investigate the nuclei with a larger number of
neutrons distributed in the halo.
For this purpose,
 probable halo phenomena have been searched for in heavier
neutron-rich nuclei~\cite{Meng-Zr}.

The Hartree-Fock-Bogoliubov (HFB)
theory~\cite{Meng-Zr,DobHFB1,DobHFB2,Bulgac,Bender2003,Meng-Li,Meng2006,Vretenar2005}
is a powerful tool to describe the heavier neutron-rich nuclei.
It can provide a unified and self-consistent
description of both the mean field and pairing
correlations in terms of the Bogoliubov quasiparticles.
Thus, the pairing properties in nuclei near the drip-line have been studied
extensively within the
relativistic Hartree-(Fock)-Bogoliubov
{scheme}~\cite{Meng-Li,Meng-Zr,Meng-Ca,MengRCHB1998,
Meng2006,Vretenar2005,ZhangSQ2003,Terasaki2006,
Zhou2003,Zhou2010,Long2010,LiLuLu2012,ChenY2012} as well as
the non-relativistic HFB scheme~\cite{DobHFB2,Bennaceur00,DD-Dob,
Bender2003,Grasso,GrassoGhalo,MMS05,Yamagami05,
Hamamoto03,Hamamoto04,Oba,ZhangY2011,Hagino11,Terasaki2006}.

The giant neutron halo with more than two neutrons is predicted first by
the relativistic continuum Hartree-Bogoliubov (RCHB)
theory~\cite{MengRCHB1998} for neutron-rich Zr~\cite{Meng-Zr,ZhangSQ2003}
and Ca~\cite{Meng-Ca,ZhangSQ2003,Terasaki2006} isotopes.
It is found that the radius of the neutron density distribution shows an
abnormal increase for $A>60$ in Ca and $A>122$ in Zr isotopes. It has been shown later
that the non-relativistic Skyrme HFB model can also describe the giant
neutron halo in these elements as far as appropriate parameter sets
are chosen~\cite{GrassoGhalo,Terasaki2006}. Recently, the giant
halo is also predicted in Ce isotopes by the relativistic
Hartree-Fock-Bogoliubov (RHFB) theory~\cite{Long2010Ce}.

In the preceding works, focuses are often put on the
extended tail of the neutron densities which leads to the abnormal
increase of the root-mean-square
(r.m.s.) radius of a halo nucleus.
It has been discussed that
the pair correlation can produce the halo
tail of the particle density via the
continuum coupling~\cite{DobHFB2,Meng-Li,Meng-Zr,Meng-Ca}
and also suppress the asymptotic particle density distribution due to
the additional binding~\cite{Bennaceur00,GrassoGhalo,Yamagami05,Hagino11}.
On the other hand, properties of
the pair density or the pairing tensor, which represents the condensate of the
nucleon pair, especially those in the low-density halo region, have not been studied
in detail.  In this paper, we would like to investigate how
 the neutron pair density behaves in
the asymptotic halo region,
and also what mechanisms govern its asymptotic behaviors.

Useful information in analyzing the above properties is the spectrum of
the neutron quasiparticle states.
Because of the shallow Fermi energy, most of the
quasiparticle states are embedded in the continuum energy region by a coupling to the
scattering states via the pair potential~\cite{DobHFB1,DobHFB2,Bulgac,Belyaev}.
In the standard HFB
calculations~\cite{Meng-Zr,DobHFB1,DobHFB2,Bulgac,Bender2003,Meng-Li,Meng2006,
Vretenar2005,Bennaceur00,DD-Dob,Yamagami05,Hagino11,MengRCHB1998,
Meng-Ca,Zhou2003,Zhou2010,LiLuLu2012,ChenY2012,Long2010,ZhangSQ2003,Terasaki2006},
a finite box or a harmonic oscillator/Woods-Saxon basis is adopted.
In this case
the continuum quasiparticle states are all discretized  (hereafter
{referred to as} the discretized HFB approach),
making it difficult to describe the asymptotic behavior and to
quantify the spectra in terms of, e.g., the resonance energy and the width.
There exist techniques to overcome this problem, for instance,  using a
very large box \cite{Yamagami05,Bennaceur00},
or adopting the analytical continuation
in the coupling constant method~\cite{YangSC,ZhangSS},
or the stabilization method~\cite{Zhang2008,Pei2011}, or
Gamow HFB approach~\cite{Michel2008}. In the present
work, we adopt a new formulation of the Skyrme HFB model
in which the Green's function method~\cite{Belyaev}
is adopted to describe
precisely the asymptotic behavior of
scattering waves for the unbound quasiparticle states in the continuum
(hereafter {referred to as} the continuum HFB approach)~
\cite{Oba,ZhangY2011}.

In this paper
we will apply this  self-consistent continuum Skyrme
HFB approach with Green's function method~\cite{ZhangY2011} to study the pairing
properties in the giant halo Zr nuclei.
In section II, we will briefly describe the formulation of
the continuum Skyrme HFB theory with the Green's function
method and the numerical details. We will also examine the significance of the
continuum in the description of the pairing properties
by comparing with the results {obtained by} the discretized HFB approach.
After presenting the results and discussions in Sections III-V
we draw conclusions in Section VI.

%
\section{\label{sec:form}Formalism}
%

The fundamental building block of the
 the Hartree-Fock-Bogoliubov (HFB) theory is the quasiparticle states. The energy $E$ and the
 wave function $\phi_i(\br\sig)$  of a quasiparticle state obeys the HFB equation
  \beq
      \int d\br'\sum_{\sig'}
      \left(
        \begin{array}{cc}
          h(\br\sig,\br'\sig')-\lam\del(\br-\br')\del_{\sig\sig'}
        & \tilde{h}(\br\sig,\br'\sig') \\
          \tilde{h}^*(\br\tilde{\sig},\br'\tilde{\sig}')
        & -h^*(\br\tilde{\sig},\br'\tilde{\sig}')+\lam\del(\br-\br')\del_{\sig\sig'} \\
        \end{array}
      \right)\phi_i(\br'\sig') = E_i \phi_i(\br\sig),
      \label{eq:HFB-1}
  \eeq
  where $\lam$ is the Fermi energy.  The Hartree-Fock (HF) Hamiltonian
  $h$ and the pair Hamiltonian $\tilde{h}$ can be obtained by
  the variation of the total energy functional with respect to
  the particle density matrix
$\rho(\br\sig, \br'\sig')=\bra\psi^\dagger(\br'\sig')\psi(\br\sig)\ket$ and
  pair density matrix
$\tilde{\rho}(\br\sig, \br'\sig')=\bra\psi(\br'\tilde{\sig}')\psi(\br\sig)\ket$, respectively.

  For the spherical system, the quasiparticle wave function
  can be written as
  \beq
    \phi_i(\br\sig) = \frac{1}{r} \phi_{lj}(r) Y_{ljm}(\hatr\sig),~~
    \mbox{where}~~ \phi_{lj}(r) = \left(
                                    \begin{array}{c}
                                     \vph_{1,lj}(r) \\
                                     \vph_{2,lj}(r)
                                    \end{array}
                                  \right).
  \eeq
  The local particle density
  ${\rho(\br)}
  =\sum_\sig\rho(\br\sig,\br\sig)$
and the pair density
  ${\tilde{\rho}(\br)}=\sum_\sig\tilde{\rho}(\br\sig,\br\sig)$
  consist of the products of the quasiparticle wave functions
summed up over all the states. Using the Green's function technique,
the local densities are expressed~\cite{Belyaev,Oba,ZhangY2011} as
  \begin{subequations}
    \beqn
      \rho(r)
       &=& \frac{1}{4\pi r^2}\sum_{lj} (2j+1)\frac{1}{2\pi i}\oint_{C_E} dE~
       {\mathcal{G}^{11}_{0,lj}(r,r,E)},  \\
  \tilde{\rho}(r)
       &=& \frac{1}{4\pi r^2}\sum_{lj} (2j+1)\frac{1}{2\pi i}\oint_{C_E} dE~
       {\mathcal{G}^{12}_{0,lj}(r,r,E)},
    \eeqn
    \label{eq:density-total}
  \end{subequations}
where  $\mathcal{G}^{11}_{0,lj}(r,r,E)$ and $\mathcal{G}^{12}_{0,lj}(r,r,E)$ are the radial HFB Green's functions.
   The Green's functions  are  constructed from the independent solutions of the radial HFB equation (\ref{eq:HFB-1}) with
  proper boundary conditions for the wave function $\phi_{lj}(r,E)$ of the quasiparticle
  states with energy $E$. Similarly, one can express other local densities needed
  in the Skyrme functional, such as kinetic energy density,
  spin-orbit densities, etc., in terms of the Green's functions.
  The integrals in Eqs.~(\ref{eq:density-total}) are
contour integrals in the complex $E$ plane, and the integration path $C_E$
  is chosen to be a rectangle with
  a height {$\gamma=0.1$}~MeV and a length $E_{\text{cut}}=70$~MeV,
  which symmetrically encloses the real negative quasiparticle energy axis
  as in Ref.~\cite{ZhangY2011}.
  The energy step of the contour integration is $\Delta E=0.01$~MeV.
  We choose the box size $R_{\rm box}=20$~fm, and
  the mesh size $\Delta r=0.2$~fm
  for the Runge-Kutta algorithm to obtain the independent
  solution of the radial HFB equation.
  The quasiparticle wave functions are connected at $r=R_{\rm box}$ to the
  asymptotic wave
  ${\phi^{\rm (out)}_{lj}(r,E)/r}= \left(A h_l^{(+)}(k_{+} r), B h_l^{(+)}(k_{-}r)\right)^{T}$
where $h_l^{(+)}(z)$ is the spherical Hankel function~\cite{Messiah} and $k_{\pm}(E)=\sqrt{2m(\lam \pm E)/\hbar^2}$.
  Other details can be found in Ref.~\cite{ZhangY2011}.

%
We employ the Skyrme functional SkI4~\cite{SkI4}, following
 Ref.~\cite{GrassoGhalo}, with which the giant halo
phenomenon in the Zr isotopes can be reproduced as predicted in the
RCHB theory~\cite{Meng-Zr}.
For the pairing interaction, we adopt the density-dependent delta
interaction (DDDI).
The difference between the zero range and finite range pairing interaction for
exotic neutron rich nucleus has been discussed in Ref.~\cite{Meng1998PRC}.
The pair field here is taken as
 \beq
   \Del(\br) = \half V_0\ls 1-\eta\lb\frac{\rho_{q}(\br)}{\rho_0}\rb^{\alp}\rs \tilde{\rho}(\br),~~q=n~{\rm or}~p.
   \label{eq:DDDI}
 \eeq
 The parameters in Eq.~(\ref{eq:DDDI}) are adopted as
 $V_0=-458.4$~MeV~fm$^{-3}$, $\eta=0.71$, $\alp=0.59$,
 and $\rho_0=0.08$~fm$^{-3}$~\cite{Matsuo2006PRC,Matsuo2007NPAc,Matsuo2010PRC,Shimoyama2011},
 which reproduce the experimental neutron pairing gap along the Sn isotopic chain.
 In particular, the parameter $V_0$ is chosen in such a way that the DDDI
 reproduces the $^{1}S$ scattering length $a=-18.5$ fm of the bare nuclear force
 in the low density limit $\rho(r) \rightarrow 0$, i.e., in the free space outside
 the nucleus.
 The  quasiparticle states are truncated at the maximal
 angular momentum $j_{\rm max}=15/2$ and at the maximal quasiparticle
 energy $E_{\text{cut}}=70$~MeV~\cite{GrassoGhalo}.

 For the sake of comparison, we also perform a box-discretized HFB calculation. In this case
  the HFB equation (\ref{eq:HFB-1}) is solved with the box boundary condition, i.e. by assuming that
  the wave functions of the quasiparticles vanish at the box boundary
  $r=R_{\rm box}$.
  The discretized quasiparticle wave functions thus obtained, $\phi_{nlj}(\br\sig)$,
  are summed up to construct the densities as
  \begin{subequations}
    \beqn
      \rho(r)
       &=& \frac{1}{4\pi r^2}\sum_{lj} (2j+1)
\sum_n {\vph^2_{2,nlj}(r)},
  \\
  \tilde{\rho}(r)
       &=& \frac{1}{4\pi r^2}\sum_{lj} (2j+1)
\sum_n {\vph_{1,nlj}(r)\vph_{2,nlj}(r)}.
    \eeqn
    \label{eq:density-box}
  \end{subequations}

%

%

\section{\label{subsec:gianthalo}Pair correlation in giant neutron halo nucleus}

\begin{figure}[!h]
\includegraphics[width=8cm]{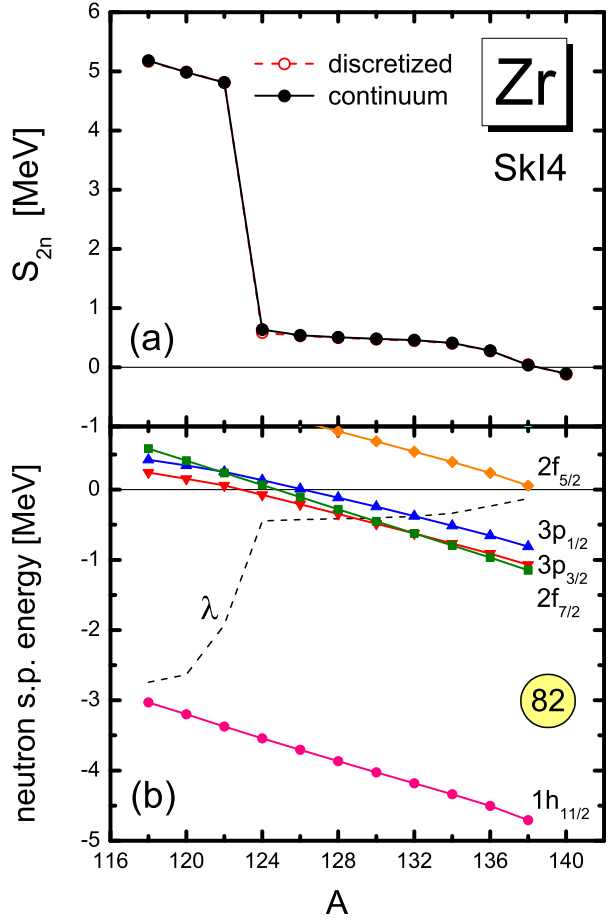}\\
\caption{(a) Two neutron separation energy $S_{2n}$ obtained
in the continuum (filled circle) and box-discretized (open circle)
Skyrme Hartree-Fock-Bogoliubov (HFB) calculations for
neutron-rich Zr isotopes with $A=118-138$.
(b) Neutron Hartree-Fock (HF) single-particle energy $\vep$ of Zr isotopes around the Fermi energy,
 which is given by the eigen solution of HF Hamiltonian $h$ after the
 final convergence of the continuum HFB calculation.
 The dashed line denotes the Fermi energy $\lam$.
  }\label{fig:Zr-S2n-spspectrum}
\end{figure}

The filled circles in Fig.~\ref{fig:Zr-S2n-spspectrum}(a) are
the two-neutron separation energy $S_{2n}(N,Z)=E(N,Z)-E(N-2,Z)$, where
$E(N,Z)$ is the total binding energy
of the isotope with $N$ neutrons and $Z$ protons obtained
in the continuum HFB calculation.
It suddenly drops from ${S_{2n}= 4.81}$ MeV at $^{122}$Zr
to $S_{2n}={0.64}$~MeV at $^{124}$Zr as the neutron number
exceeds the magic number $N=82$.
Then it gradually decreases  to an extremely small value
${0.04}$~MeV at $^{138}$Zr,
and finally becomes negative in $^{140}$Zr.
Thus $^{138}$Zr is the neutron drip-line nucleus in the present model,
which is consistent with the previous investigation~\cite{GrassoGhalo}.
The small
two-neutron separation energy is one of the conditions for
the emergence of the
halo structure.
Figure~\ref{fig:Zr-S2n-spspectrum} (b) shows the neutron Fermi energy $\lambda$ as well as the
Hartree-Fock (HF) single-particle energies $\vep$,
which are the eigen energies
of HF Hamiltonian $h$ (obtained after the final convergence
of the continuum HFB calculation).
One can see that, as the neutron number increases,
the Fermi energy is raised up, to a position  quite close to the continuum threshold,
while all the HF single-particle orbits fall down.
Explicitly, the $3p_{1/2}$, $3p_{3/2}$ and $2f_{7/2}$ states evolve from
unbound resonances ($A \leq 122$) to weakly bound orbits ($A\gtrsim 126$), and
the $2f_{5/2}$ orbit remains as a resonance in the continuum.
The values of these HF single-particle energies in $^{124}$Zr
 $\sim ^{138}$Zr can be found in Table~\ref{tab:Egam-all}.
The HF Hamiltonian has another resonance $1h_{9/2}$
which lies slightly higher than the plotted energy region in Fig.~\ref{fig:Zr-S2n-spspectrum}~(b).

\begin{figure}[!h]
\includegraphics[width=12cm]{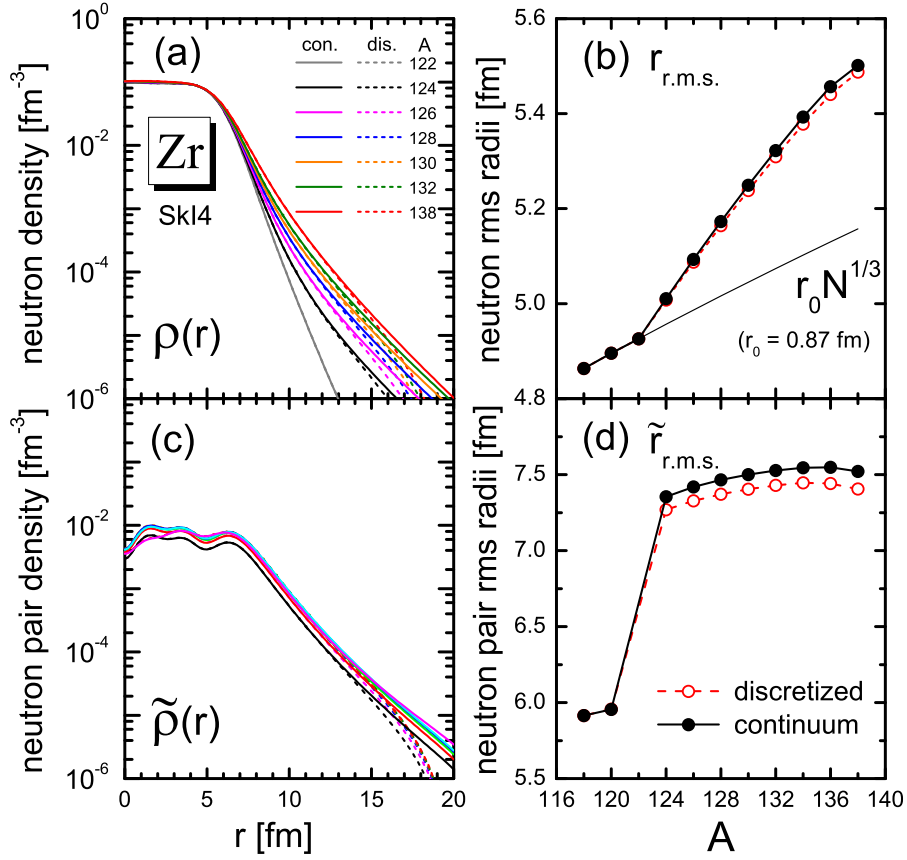}
\caption{(a) Neutron density $\rho(r)$, (b)
{neutron} root-mean-square (r.m.s.) radius $ r_{\rm r.m.s.}$,
(c) neutron pair density $\tilde{\rho}(r)$ and
(d) neutron {pair} r.m.s. radius
$\tilde{r}_{\rm r.m.s.}$ for Zr isotopes.
The solid lines and the filled circles are the results of
the continuum HFB calculation while the dotted lines and
the open circles are those obtained in the box-discretized HFB
calculation.
}
 \label{fig:Zr-density-rms}
\end{figure}

Figure~\ref{fig:Zr-density-rms}~(a) shows the neutron density $\rho(r)$
calculated for the Zr isotopes
with $A=122-138$ using the continuum HFB calculation (solid lines).
It is seen that the neutron
density for $A=124-138$ exhibits a long tail extending far outside
the nuclear surface compared with $A=122$.
Figure~\ref{fig:Zr-density-rms}~(b) shows the
r.m.s.
radius $r_{{\rm r.m.s.}} \equiv \ls \int 4\pi r^4\rho(r)dr/
\int 4\pi r^2\rho(r)dr \rs^{1/2}$ calculated for the corresponding neutron density
(filled circles).
Compared with the isotopic trend in $A\le 122$,
which gives an extrapolation as $r_{{\rm r.m.s.}} \approx 0.87 N^{1/3}$~fm,
the neutron r.m.s. radius
in $^{124}$Zr and
the heavier isotopes
displays a steep increase with $N$. These results are
consistent with the previous investigations in
Refs.~\cite{Meng-Zr} and~\cite{GrassoGhalo},
where the giant halo structure, i.e. the long tail which can
accommodate more than two neutrons, is predicted
in the isotopes with $A\ge 124$.

Figure~\ref{fig:Zr-density-rms}~(c) shows the neutron pair density $\tilde{\rho}(r)$.  We clearly see that the
pair density has an even more significant tail, which is characterized by i) $\tilde{\rho}(r) > \rho(r)$
for ${r \gesim 10}$ fm, and ii) that $\tilde{\rho}(r)$ has a more gentle slope than $\rho(r)$. These features
are reflected in the r.m.s. radius weighted with the
neutron pair density,
$\tilde{r}_{{\rm r.m.s.}} \equiv \ls \int 4\pi r^4\tilde{\rho}(r)dr/
\int 4\pi r^2\tilde{\rho}(r)dr \rs^{1/2}$, which is
plotted in Fig.~\ref{fig:Zr-density-rms}~(d).
{We shall call $\tilde{r}_{{\rm r.m.s.}}$ the pair r.m.s. radius hereafter.}
It is significantly larger than
$r_{{\rm r.m.s.}}$ of the neutron density. Also,  the
neutron pair r.m.s. radius $\tilde{r}_{{\rm r.m.s.}}$ exhibits
a sudden jump at $^{124}$Zr, and remains almost the same
until the drip-line.  Here,  $\tilde{r}_{{\rm r.m.s.}}$ of
$^{122}$Zr is omitted in the plot due to the absence of pairing.

\begin{figure}[!h]
\includegraphics[width=12cm]{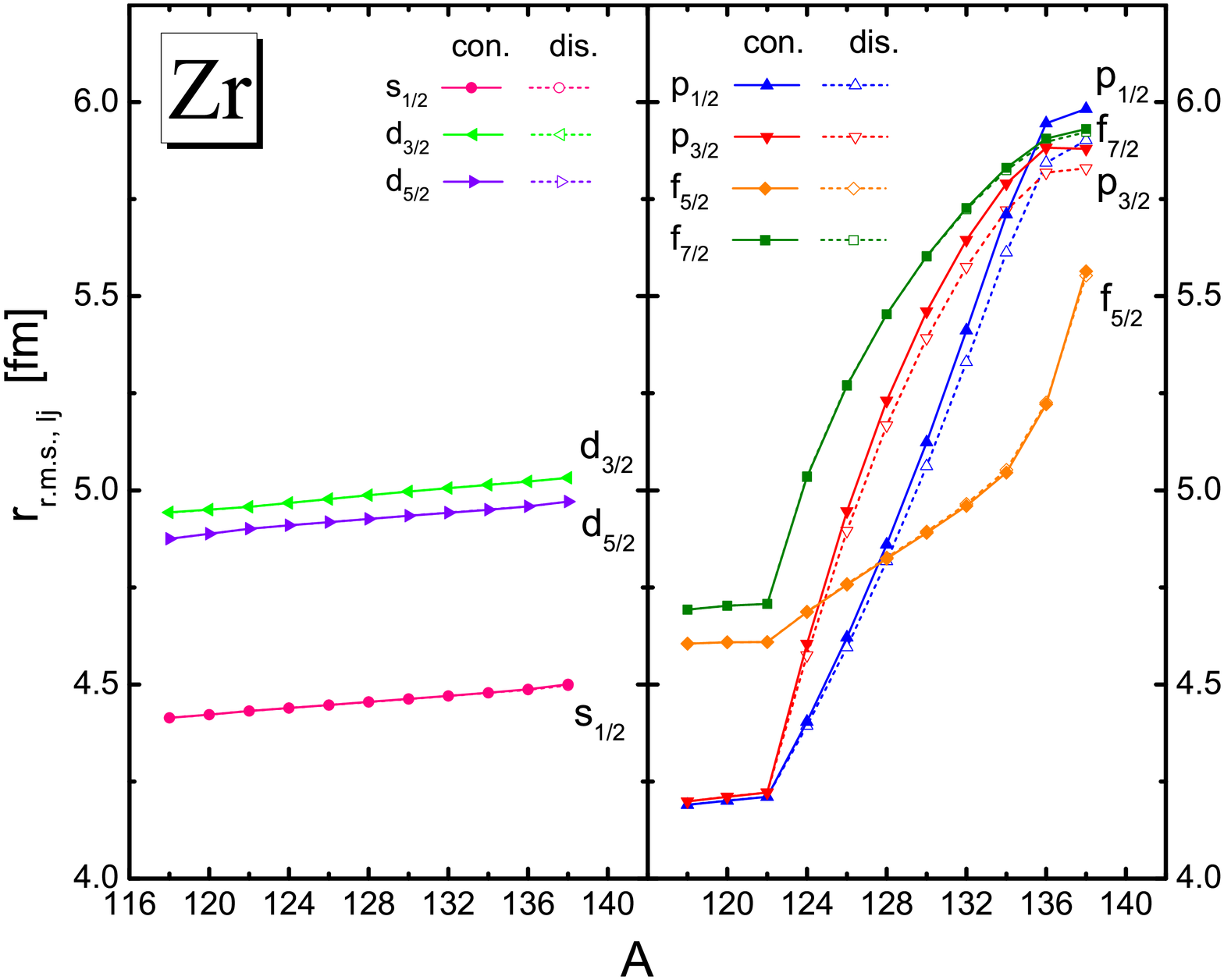}
\caption{{Neutron} root-mean-square radius $r_{{\rm r.m.s., }lj}$ of the $s$, $p$, $d$ and $f$ partial waves of Zr isotopes calculated for the $lj-$decomposed neutron density ${\rho}_{lj}(r)$.
The filled symbols are the results obtained in the continuum HFB calculation
while the open symbols are the results obtained in the box-discretized HFB calculation.
  }\label{fig:Zr-rms-spdf}
\end{figure}

It is argued in Refs.~\cite{Meng-Zr} and~\cite{GrassoGhalo}
that the neutron orbits $3p_{1/2}, 3p_{3/2}, 2f_{5/2}$ and $2f_{7/2}$
around the Fermi energy
play a decisive role in forming the giant halo structure
in the neutron density.
Similarly to Refs.~\cite{Meng-Zr} and~\cite{GrassoGhalo},
we decompose the neutron density with respect to different
partial waves $lj$, and calculate separately the r.m.s. radius
 \beq
    r_{{\rm r.m.s.}, lj} =\lb \frac{\int 4\pi r^4 \rho_{lj}(r) dr}
  {\int 4\pi r^2 \rho_{lj}(r) dr}\rb^{1/2}
 \eeq
weighted by the corresponding $lj$-decomposed neutron density
  \beqn
   \rho_{lj}(r)
       = \frac{(2j+1)}{4\pi r^2}\frac{1}{2\pi i}\oint_{C_E} dE~
       {\mathcal{G}^{(11)}_{0,lj}(r,r,E)},
       \label{eq:rhojl-GF}
  \eeqn
 which sums the contributions from the quasiparticles with the quantum
 number $lj$ within $E=0\sim E_{\rm cut}$.
 The  r.m.s. radii for
 $lj=s_{1/2},~p_{1/2},~p_{3/2},~d_{3/2},~d_{5/2},~f_{5/2}$ and
 $f_{7/2}$ partial waves are plotted (filled symbols) in Fig.~\ref{fig:Zr-rms-spdf}.
 We can see clearly that
 the increase of the total r.m.s. radius is mainly
 contributed by the $p$ and $f$ states, in agreement with  Refs.~\cite{Meng-Zr,GrassoGhalo}.

\begin{figure}[!h]
\includegraphics[width=12cm]{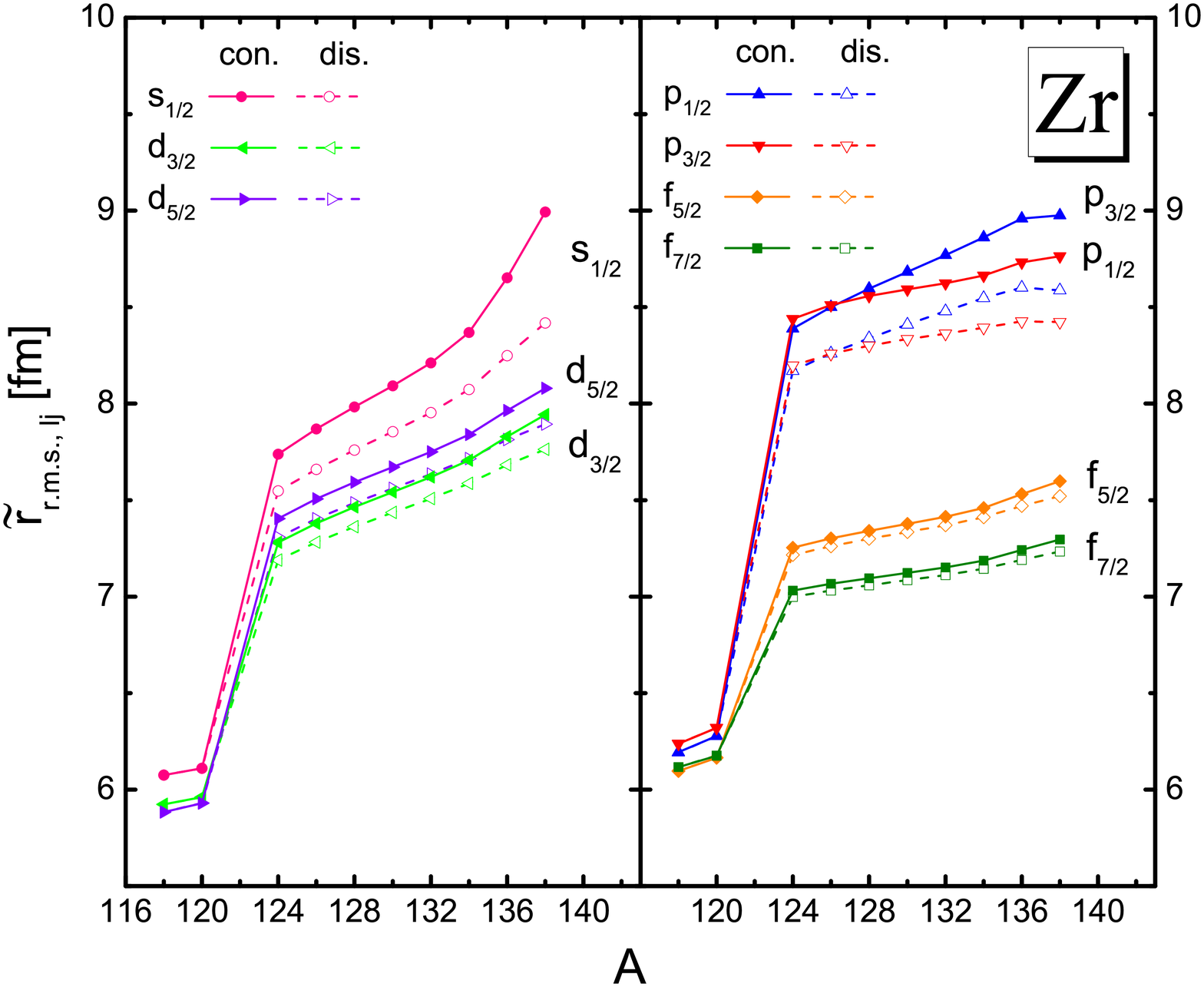}
\caption{
{Neutron} pair root-mean-square radius $\tilde{ r }_{{\rm r.m.s., }lj}$ of the $s$, $p$, $d$ and $f$ partial waves of Zr isotopes calculated for the $lj$-decomposed neutron pair density $\tilde{\rho}_{lj}(r)$.
The filled symbols are the results obtained in the continuum HFB calculation
while the open symbols are the results obtained in the box-discretized HFB calculation.
  }\label{fig:Zr-rmstjl-spdf}
\end{figure}

 The pair density displays an apparently different behavior
 from that of the neutron density as seen in Fig.~\ref{fig:Zr-rmstjl-spdf}, in which
 we plot the neutron pair r.m.s. radius
 \beq
    \tilde{r}_{{\rm r.m.s.}, lj} =\lb \frac{\int 4\pi r^4 \tilde{\rho}_{lj}(r) dr}
  {\int 4\pi r^2 \tilde{\rho}_{lj}(r) dr}\rb^{1/2},
  \label{eq:rhotjl-GF}
 \eeq
 weighted by the $lj$-decomposed neutron pair density
  \beqn
   \tilde{\rho}_{lj}(r)
       = \frac{(2j+1)}{4\pi r^2}\frac{1}{2\pi i}\oint_{C_E} dE~
       {\mathcal{G}^{(12)}_{0,lj}(r,r,E)}.
  \eeqn
 A notable feature is that the jump
 at $N=124$ is seen not only for the partial waves
 $p_{1/2}, p_{3/2}, f_{5/2}$ and $f_{7/2}$, which have HF
 single-particle orbits or resonances near the Fermi energy,
 but also for the partial waves $s_{1/2}, d_{3/2}$ and $d_{5/2}$.
 This indicates that the $s$- and $d$-waves
 also play important roles in the large tail of the pair density.

\begin{figure}[!h]
\includegraphics[width=12cm]{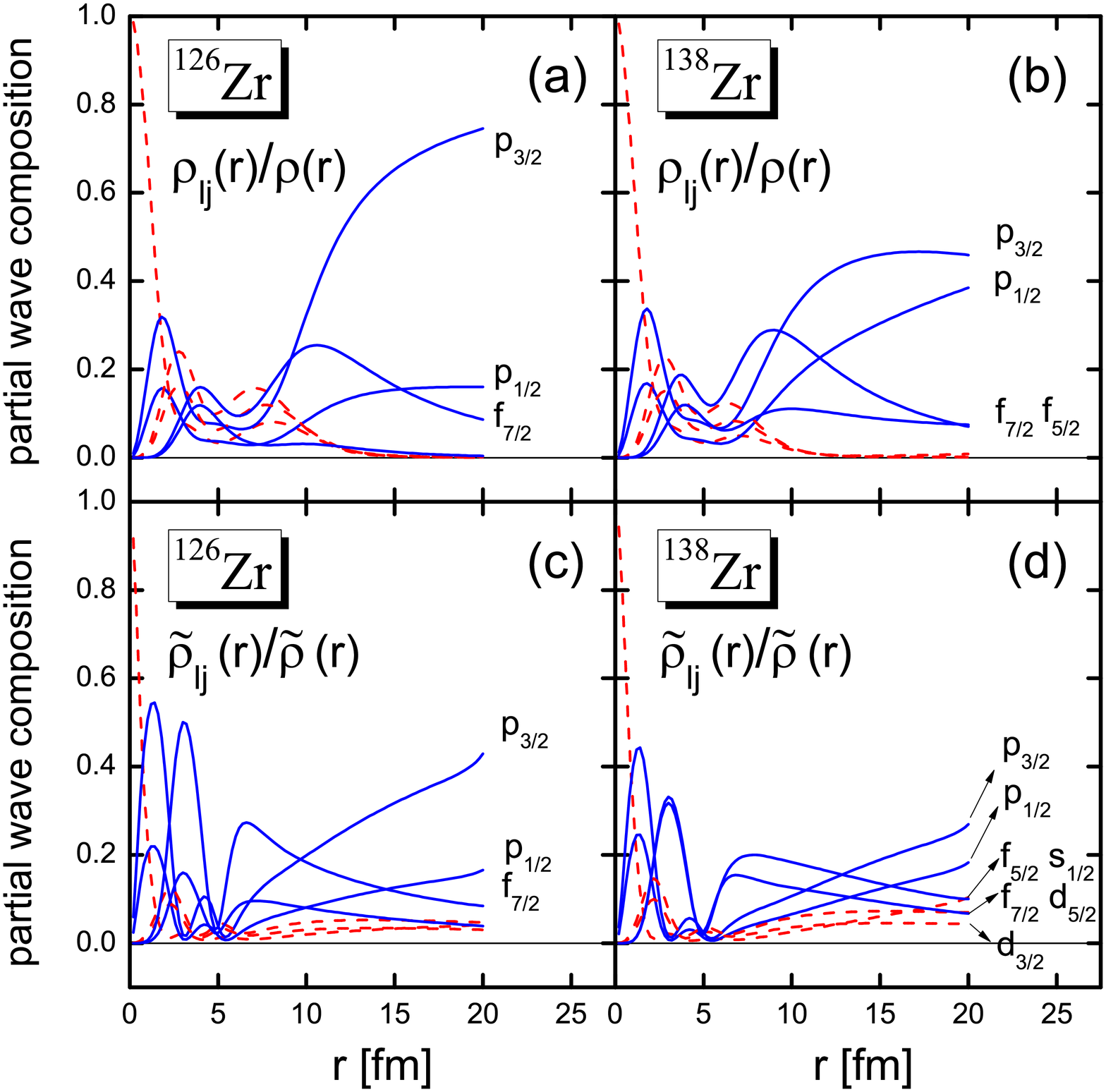}
\parbox{16cm}{
\caption{Compositions of different partial waves to the total neutron density,
$\rho_{lj}(r)/\rho(r)$, as a function of the radial coordinate $r$ for
$^{126}$Zr (panel (a)) and $^{138}$Zr (panel (b)).
The solid lines denote the negative parity states, and the dashed lines
the positive parity states.  The same plot for the neutron pair density,
$\tilde{\rho}_{lj}(r)/\tilde{\rho}(r)$, for
$^{126}$Zr (panel (c)) and $^{138}$Zr (panel (d)).
  }\label{fig:Zr-density-ratio}
  }
\end{figure}

 To clarify the difference,  we present  in Fig.~\ref{fig:Zr-density-ratio} the composition of the neutron density
 ${\rho}(r)$ and pair density $\tilde{\rho}(r)$ with respect to the partial waves $lj$, i.e., $\rho_{lj}(r)/\rho(r)$ and
 $\tilde{\rho}_{lj}(r)/\tilde{\rho}(r)$ for $^{126}$Zr and $^{138}$Zr.
 As shown in the previous works~\cite{Meng-Zr}
 and~\cite{GrassoGhalo},
 the most dominant compositions
of the neutron density $\rho(r)$ in the halo region
${r\gesim 10}$~fm are
 the $p$-waves, and the
 next are the $f$-waves, while the
 $s$- and $d$-waves as well as others partial waves have
 very little contributions, and almost vanish at
 large distances ${r\gesim 15}$ fm.

 For the pair density $\tilde{\rho}(r)$,
 however, we can see from Fig.5 (c) and (d)
 that the partial wave composition is
 very different from that of the neutron density $\rho(r)$.
 The dominance of the waves $p_{1/2}$, $p_{3/2}$, $f_{5/2}$ and $f_{7/2}$
 is commonly seen also for the pair density, but to a less extent.
 The waves  $s_{1/2}, d_{3/2}$ and $d_{5/2}$, on the other hand,
 have small but non-negligible contributions in the halo region
 ${r \gesim 10}$ fm. This feature becomes even stronger in $^{138}$Zr
 than in $^{126}$Zr. Since there is no bound HF single-particle orbits
 nor resonance orbits in the $s$- and $d$-waves near the Fermi energy,
 it is non-resonant continuum orbits that contribute to the pair density.

Before closing this section, we would like to compare
the present results obtained in the
continuum HFB calculation with those obtained in
the box-discretized HFB calculation.  In the latter case, the quasiparticle states are discretized
even in the continuum energy region $E>|\lambda|$,
and the densities are obtained as Eq.~(\ref{eq:density-box}).

The two-neutron separation energy
$S_{2n}$ in this approximation is shown
in Fig.~\ref{fig:Zr-S2n-spspectrum}~(a) with
the open circles.
The total neutron density and pair densities of the
box-discretized results are shown with the dotted
lines in Fig.~\ref{fig:Zr-density-rms}~(a) and (c).
The corresponding neutron r.m.s. radii
$r_{{\rm r.m.s.}}$ and pair r.m.s. radii $\tilde{r}_{{\rm r.m.s.}}$ are shown
with open circles in panels (b) and (d).
The r.m.s. radii of different partial waves are also shown
in Figs.~\ref{fig:Zr-rms-spdf} and~\ref{fig:Zr-rmstjl-spdf}
with open symbols for the neutron density and the pair density, respectively.

From the comparison between the continuum and the box-discretized
calculations, we observe the following.
The influence of the box boundary condition
is seen both in the neutron density $\rho(r)$ and in the pair density
$\tilde{\rho}(r)$ at
$r\approx 15 - 20$ fm close to the box boundary
(Fig.~\ref{fig:Zr-density-rms}~(a) and (c)).
However this influence causes only negligibly small difference
in the bulk properties such as the total energy and its derivative,
the two-neutron separation energy $S_{2n}$.
This is because the density and
the pair density of neutrons at $r \gesim 15$ fm are very
small, and its contribution to the total energy is accordingly small.
The r.m.s. radius $r_{{\rm r.m.s.}}$ (Fig.~\ref{fig:Zr-density-rms}~(b))
of the total neutron density is also affected very little by the
box-discretization.
If we look into the r.m.s. radii $r_{{\rm r.m.s.}, lj}$ for each partial waves (Fig.~\ref{fig:Zr-rms-spdf}), we can see some difference in
the contributions from the $p$-wave while the
other partial waves are not affected.
This indicates that the neutron density contributed from the $p$-wave
is the most extended while
other partial waves extend less (see also Fig.\ref{fig:Zr-density-ratio}).
More importantly, we can see more significant difference
in the neutron pair r.m.s. radii $\tilde{r}_{{\rm r.m.s.}}$ and
$\tilde{r}_{{\rm r.m.s.},lj}$
(cf. Fig.~\ref{fig:Zr-density-rms}~(d)
and Fig.~\ref{fig:Zr-rmstjl-spdf}). The difference is
seen not only in the $p$- and $f$-waves, but also in
the $s$- and $d$-waves. This difference is larger
 for lower angular momentum states and in
 more neutron-rich nuclei, where we expect more
 continuum coupling due to the lower centrifugal barrier.
 It can be concluded from these observations that
the proper treatment of the continuum quasiparticle states
is important for the description of the tail part of the pair density,
i.e. the pair correlation in the giant halo.

\section{Continuum quasiparticle spectra}

\subsection{Low-lying quasiparticle states}

A quasiparticle state with excitation energy $E$ in the partial wave $lj$ has
contributions to the density $\rho(r)$ and the pair density $\tilde{\rho}(r)$, which are
given by~\cite{Oba,ZhangY2011}
\beqn
  {\rho}_{lj}(r,E) &=&\frac{(2j+1)}{4\pi r^2}\frac{1}{\pi}\text{Im}
   \mathcal{G}_{0,lj}^{(11)}(r,r,-E-i\eps),  \cr
  \tilde{\rho}_{lj}(r,E) &=&\frac{(2j+1)}{4\pi r^2}\frac{1}{\pi}\text{Im}
   \mathcal{G}_{0,lj}^{(12)}(r,r,-E-i\eps),
\label{eq:qpspectrum-rhore}
 \eeqn
 with the use of the HFB
 Green's function and an infinitesimal  constant $\eps$.
 They satisfy  $\rho(r)=\sum_{lj}\int dE {\rho}_{lj}(r,E) $ and
 $\tilde{\rho}(r)=\sum_{lj}\int dE \tilde{\rho}_{lj}(r,E). $ We can then
define
the occupation number density as a function of $E$ by
 \beq
  {n}_{lj}(E) = \int 4\pi r^2{\rho}_{lj}(r,E)dr,
 \eeq
 which
 represents the neutron number associated with the quasiparticle
state with energy $E$ and the quantum numbers $lj$.
 We can also define
the pair number density
 \beq
  \tilde{n}_{lj}(E) = \int 4\pi r^2\tilde{\rho}_{lj}(r,E)dr,
 \eeq
using the pair density contribution $\tilde{\rho}_{lj}(r,E)$.

\begin{figure}[!h]
\includegraphics[width=18cm]{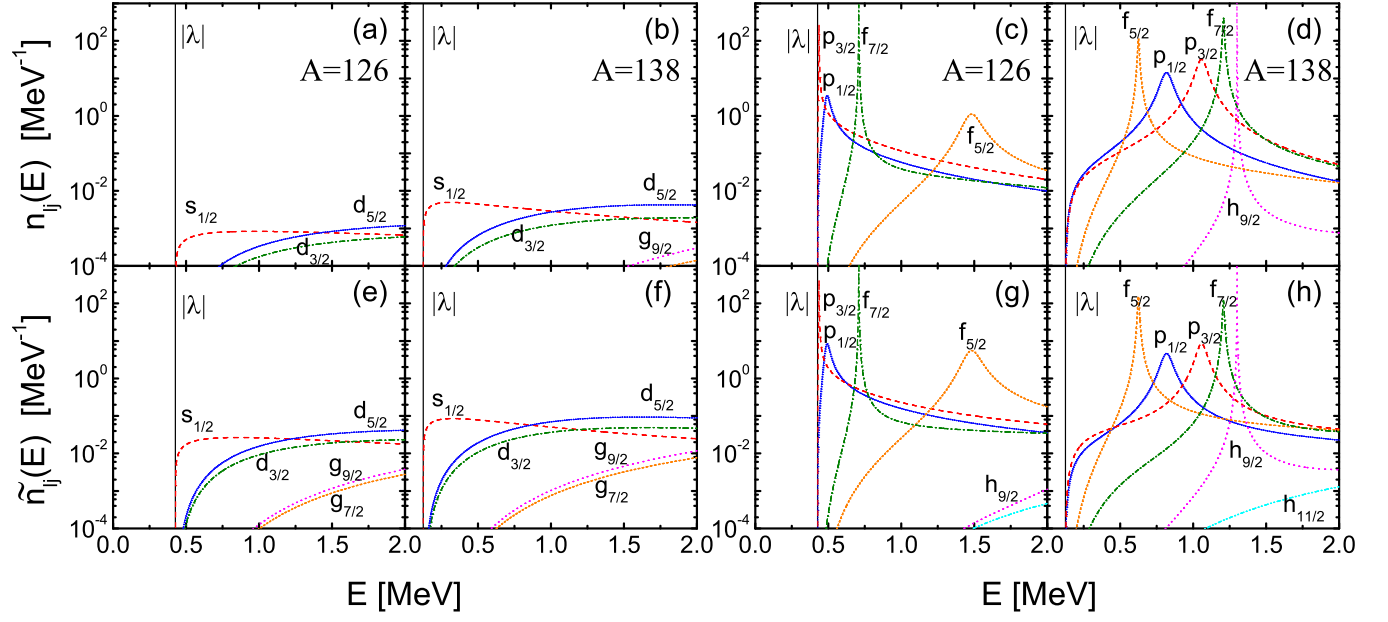}
\vspace{-0.5cm}
\caption{Occupation number densities $n_{lj}(E)$ (panels (a) and (c))
and pair number densities $\tilde{n}_{lj}(E)$ (panels (e) and (g)) of neutron
quasiparticle states for different partial waves
of $^{126}$Zr, while panels (b), (d), (f), and (h) are the same
quantities but for $^{138}$Zr. The vertical
line indicates the threshold energy $|\lam|$ for the continuum quasiparticle
states.
  }\label{fig:Zr-qpspectrum-Zr126-138}
\vspace{-0.2cm}
\end{figure}

Figure~\ref{fig:Zr-qpspectrum-Zr126-138} (a)-(d) show the
 occupation number density $n_{lj}(E)$ of neutron
 quasiparticle states in a quasiparticle energy
interval $E=0\sim 2$~MeV for $^{126}$Zr and $^{138}$Zr, while
Fig.~\ref{fig:Zr-qpspectrum-Zr126-138} (e)-(h) are the neutron pair
number density $\tilde{n}_{lj}(E)$ for the same isotopes.
We choose $\eps=10^{-10}$~MeV
in Eq.~(\ref{eq:qpspectrum-rhore}).
Note that the quasiparticle states with  energy $E$ larger than
the threshold energy $|\lam|$ form a continuum spectrum.

We can see from this figure that the quasiparticle spectra of $^{126}$Zr
and those of $^{138}$Zr have common features.
The partial waves $p_{1/2}, p_{3/2}, f_{5/2}$ and $f_{7/2}$ in
panels (c), (d), (g) and (h) have
peak structures with finite width, i.e., quasiparticle resonances,
in the spectra of  ${n}_{lj}(E)$ and
$\tilde{n}_{lj}(E)$. These quasiparticle resonances
correspond to the HF single-particle orbits which are shown
 in Fig.\ref{fig:Zr-S2n-spspectrum}~(b).
It should be noticed here that in $^{126}$Zr
the HF single-particle orbits $3p_{3/2}$ and $2f_{7/2}$ are discrete bound states with
very small binding energy $\vep \approx -0.1 \sim -0.2$ MeV.
It is the pair correlation that transforms these bound HF single-particle
orbits to unbound quasiparticle resonances.
Similarly,  the weakly bound HF single-particle orbits $3p_{3/2}, 3p_{1/2}$
and $2f_{7/2}$ in $^{138}$Zr become the quasiparticle resonances
when the pair correlation is taken into account. Table 1 lists
the resonance energies $E_{\rm res}$ of these quasiparticle resonances,
which we evaluate as the peak energy of $n_{lj}(E)$.

{Another noticeable feature is seen in
the quasiparticle spectra
of the partial waves with positive parity,
$s_{1/2}, d_{3/2}, d_{5/2}$, etc, plotted in
Fig.~\ref{fig:Zr-qpspectrum-Zr126-138}~(a), (b), (e) and (f).}
We see here smooth profiles of
${n}_{lj}(E)$ and $\tilde{n}_{lj}(E)$, indicating non-resonant
continuum quasiparticle states. Concerning the
occupation number density ${n}_{lj}(E)$, the positive parity
non-resonant quasiparticle states have
 contributions which
are smaller, by several orders of magnitude,
than the contributions of the resonant quasiparticle states
with the negative parity shown in panels (c) and (d). On the other
hand, the pair number densities $\tilde{n}_{lj}(E)$ of the
positive parity partial waves ((e) and (f)) are
 more than
ten times larger than ${n}_{lj}(E)$ of the same partial waves ((a) and (b)).
It is also noticed that
 $\tilde{n}_{lj}(E)$ of
the positive parity partial waves ((e) and (f))
are even comparable to those of the negative parity partial waves ((g) and (h)),
except around the resonant peaks.
These sizable contributions from the non-resonant
$s_{1/2}, d_{3/2}$ and $d_{5/2}$ quasiparticles
to the pair number density can be related to the
non-negligible fraction $\tilde{\rho}_{lj}(r)/\tilde{\rho}(r)$
of the $s_{1/2}, d_{3/2}$ and $d_{5/2}$  quasiparticles in
the halo tail of the pair density as shown in
Fig.~\ref{fig:Zr-density-ratio} (c) and (d).

\subsection{Pairing gaps and missing bound quasiparticle states}

In the isotopes $^{126-138}$Zr, all the quasiparticle states are embedded in
the continuum energy region $E> |\lambda|$
(Fig.~\ref{fig:Zr-qpspectrum-Zr126-138} is an example),
namely there exist no bound
quasiparticle states with $E < |\lambda|$ except in $^{124}$Zr.
[In $^{124}$Zr there is only one bound quasiparticle state $3p_{3/2}$ at
$E=0.436$ MeV, which is very close to $|\lam|=0.446$ MeV.]
  It is a remarkable
 effect of the pair correlation since
  the HF single-particle states
 $3p_{1/2}, 3p_{3/2}$ and $2f_{7/2}$ are all bound orbits
 if the pair correlation were neglected~(cf.
Fig.~\ref{fig:Zr-S2n-spspectrum}(b)).
To understand
this, we first evaluate the average pairing gaps defined as usual by
$\Delta_{uv}=\int d\br \Delta(\br)\tilde{\rho}(\br)
/\int d\br\tilde{\rho}(\br)$  and
$\Delta_{vv}=\int d\br \Delta(\br){\rho}(\br)
/\int d\br{\rho}(\br)$ with two different weight factors given by
the neutron density or the neutron pair density.
They are listed in Table~\ref{tab:Egam-all}.  Their values
$\Delta_{uv} \approx 0.59-0.68$ MeV and
 $\Delta_{vv} \approx 0.53-0.61$ MeV are fairly constant
($\Delta \approx 0.5-0.7$ MeV) in the interval $A=126-138$,
 except for $\Delta \approx 0.42-0.47$ MeV at $A=124$,
 where basically only two valence neutrons participate in the pair correlation.
 We observe here that the average pairing gap $\Delta \approx 0.5-0.7$ MeV is larger than the absolute value of the Fermi energy $|\lambda|$  in the isotopes $^{126-138}$Zr, where $|\lambda| <0.42 $ MeV.
 This relation $\Delta > |\lam|$ is proposed in Ref.~\cite{DobHFB2} as a criterion for a non-perturbative role of the pairing on the
  properties of the weakly bound nuclei.
  We can also examine this criterion with respect to the individual quasiparticle resonances.  We here
note that the energy of the quasiparticle resonances $3p_{1/2}$ and $3p_{3/2}$
satisfy $E_{\rm res} >0.4$ MeV and $E_{\rm res} > |\lambda|$  in the isotopes
with  ${A=126-138}$ (and similarly,
$E_{\rm res} >0.7$ MeV and $E_{\rm res} > |\lambda|$ for
$2f_{7/2}$). As a result, if we use the BCS expression
 $E_{\rm res}\approx \sqrt{(\vep-\lambda)^2 + \Delta^2_{\rm eff}}$ to relate the quasiparticle
 energy $E_{\rm res}$ and the effective pairing gap $\Delta_{\rm eff}$
 relevant to the specific quasiparticle
state, we can estimate the effective pairing gap of the
 $3p_{1/2}$ and $3p_{3/2}$ resonances as $\Delta_{\rm eff}\gesim 0.4$ MeV
 (and  $\Delta_{\rm eff}\gesim 0.7$ MeV for the $2f_{7/2}$ resonance)  and
 $\Delta_{\rm eff}\gesim |\lambda|$.
 [ Here we used also the fact that
 the HF single-particle energies of $3p_{1/2}, 3p_{3/2}$  and $2f_{7/2}$
are almost degenerate with
 the Fermi energy, $\vep \approx \lambda$ (cf. Fig.~\ref{fig:Zr-S2n-spspectrum}).]
 Again we can infer the relation $\Delta_{\rm eff}  > |\lam|$ also for the
effective pairing gap of the weakly-bound orbits under consideration.

 \subsection{Width of quasiparticle resonances}

\begin{figure}[htb]
\parbox{16cm}{
\includegraphics[width=15cm]{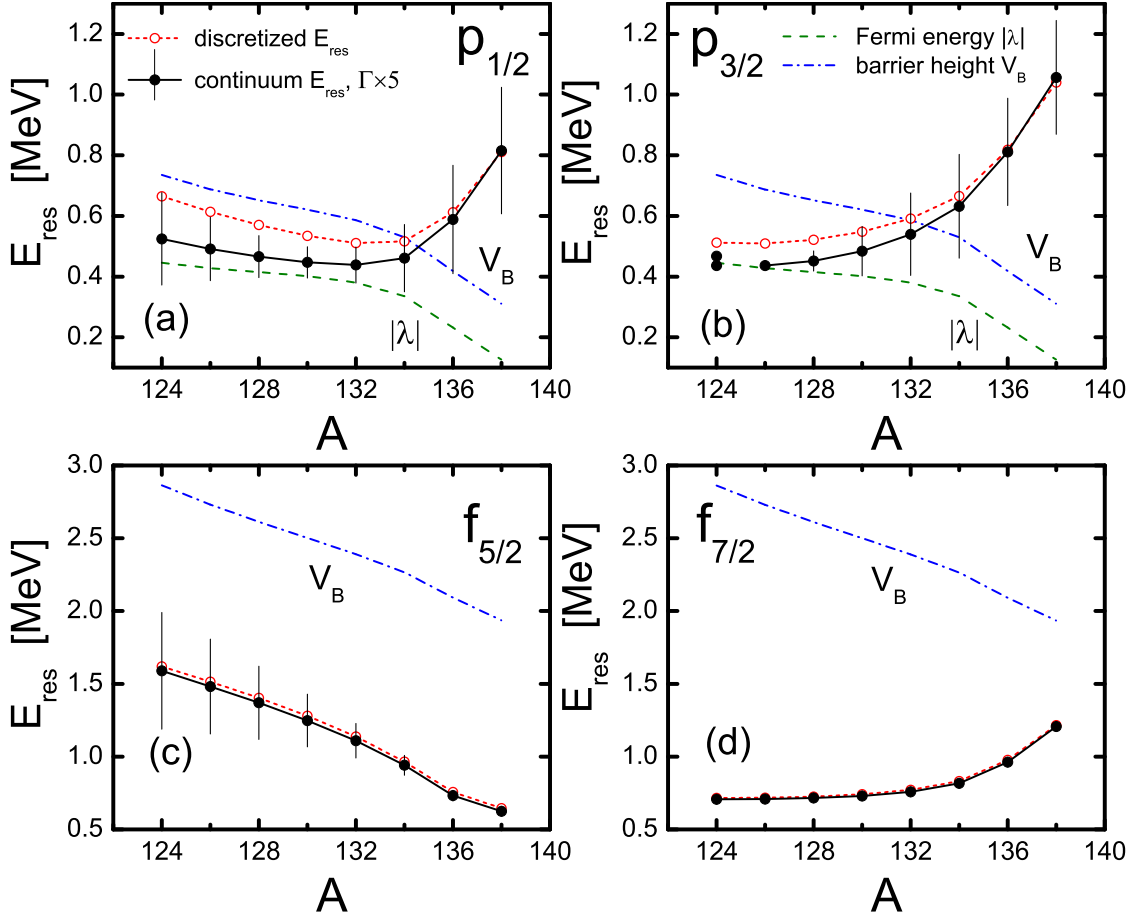}
\caption{Peak energy $E_{\rm res}$ and the width $\Gam$
of the resonant quasiparticle states around the Fermi energy in the
$p_{1/2},~p_{3/2},~f_{5/2}$, and~$f_{7/2}$ partial waves,
plotted in panels (a), (b), (c) and (d)
respectively, obtained for the Zr isotopes with $A=124-138$.
  The filled circles are
  the resonance energy $E_{\rm res}$,
  and the vertical bars represent the width $\Gamma$ multiplied by a
  factor of $5$.
  The open circles are discretized quasiparticle energies
  obtained in the box-discretized HFB calculation.
  The dashed line in panels (a) and (b) is the threshold $|\lam|$
  for the continuum quasiparticle states.
  The dash-dotted line in each panel denotes the position $V_B$
of the barrier top of the
{HF} potential
including the centrifugal potential
measured from the Fermi energy.
  }\label{fig:Zr-Egam-discon-2}
  }
\end{figure}

We evaluate the width of a quasiparticle resonance by reading the
FWHM of the peak structure in
the occupation number density $n_{lj}(E)$.
The resonance energies and widths
of the low-lying quasiparticle resonances
$3p_{1/2}, 3p_{3/2}, 2f_{5/2}$ and $2f_{7/2}$ are displayed in
Fig.~\ref{fig:Zr-Egam-discon-2}.
The length of the bar in this plot represents the width $\Gam$ of the resonant states multiplied by a factor of $5$.
For the sake of
reference, it also plots the threshold energy
$|\lambda|$ and the position $V_B$
of the barrier top of the
{HF} potential
including the centrifugal potential
measured from the Fermi energy.
We list in Table 1 the barrier height of the potential
$V_{\rm max}=V_B-|\lam|$ and the resonance
energy $e_{\rm res} = E_{\rm res} - |\lambda|=E_{\rm res}+\lambda$
measured from zero HF single-particle energy.

There are at least two mechanisms that
govern the width of the quasiparticle resonances.
One is the barrier penetration of a single-particle motion,
which is present in any potential models, i.e.
the HF potential in the present case.
The other is the one caused by the pair correlation, through which
even bound HF single-particle orbits can couple to the scattering states.
An example of the latter is known
as quasiparticle resonances originating from deep hole states
\cite{Belyaev,DobHFB2}.

These two mechanisms are interwoven in a rather complex way to produce the
widths of the low-lying quasiparticle resonances under discussion.
 A typical example is seen
in the isotopic dependence of the
width of the $3p_{1/2}$ resonance (Fig.\ref{fig:Zr-Egam-discon-2}~(a)).
As the mass number increases
from $A=124$ to $A=138$, the
width of the $3p_{1/2}$ resonance first decreases,
then turns to increase at larger neutron number $A \gesim 132$.
The decrease of the width in the interval  $124 \le A \lesim132$  is
hardly explained by the barrier penetration mechanism alone:
the barrier height $V_{\rm max}$ decreases (see Table~\ref{tab:Egam-all})
while the resonance energy $e_{\rm res}$
are fairly constant in this interval,  thus the barrier penetration
mechanism would lead to an "increase" of the width.
The decrease of the
width may be explained by combining the potential barrier penetration
and the continuum coupling caused by the pairing
in the following way. We here note that
because of the change in the relative ordering of the HF single-particle
energy $\vep$ and the Fermi energy $\lam$ (cf. Fig.2),
the particle-character of the $3p_{1/2}$ quasiparticle state, dominant at
$A=124$, is weakened gradually with increasing the mass number from
$A=124$ to $A\approx 132$, and
a hole-character grows with further increase of $A$.
 If we can assume that
 the coupling of the
  hole-component $\varphi_{2,lj}(r)$ to the scattering wave
via the pair potential is weaker than that of
the particle-component
 $\varphi_{1,lj}(r)$ to the scattering wave via the barrier penetration,
 we then expect the decrease of the width.
At the mass numbers
$A\approx 132-134$,  the barrier penetration becomes progressively
effective since the resonance energy $e_{\rm res}$ increases, and it causes the
increase of the width.  However, the width remains finite
even at $A=136,138$ where the resonance energy lies above the
barrier height. In these isotopes, the width is probably controlled
dominantly by the coupling to the continuum via the pair potential.
These speculations remain at a qualitative level.
The analytical evaluation of the width of the quasiparticle resonance is known
only for those originating from deep hole states~\cite{Belyaev,DobHFB2},
but our situation is more complex. Quantitative understanding of the width of
the low-lying quasiparticle resonances associated with the weakly-bound single-particle
orbits remains as a future subject.

\subsection{Continuum coupling and comparison with the box discretization}

It is interesting to compare the box-discretized HFB calculation
and the continuum HFB calculation for the description of
the resonant quasiparticle states.

The quasiparticle states are all discretized when the box boundary condition is adopted,
and the discretized states corresponding to the
quasiparticle resonances we are discussing
are the lowest energy states in each of the partial
waves $p_{1/2}, p_{3/2}, f_{5/2}$ and $f_{7/2}$.
The quasiparticle energies of these states are shown
in Fig.\ref{fig:Zr-Egam-discon-2} with open circles.
We see large deviations (the deviation
is even larger than the value of the resonance width) from the resonance energies
$E_{\rm res}$ obtained in the continuum HFB calculation
for the $3p_{1/2}$ and $3p_{3/2}$ states
at $A\lesim 134$. Contrastingly, at $A=136 {\rm~and~} 138$, the quasiparticle
energies obtained in the box-discretized calculation agree rather
well with the resonance energies (the deviation is comparable
with the value of the resonance width). A good agreement is also
seen in the quasiparticle resonances of $2f_{7/2}$ and $2f_{5/2}$.

A hint of the large deviation is suggested in the ratio between the width $\Gamma$
and the resonance energy $e_{\rm res}$. It is seen from Table~\ref{tab:Egam-all} that
the quasiparticle resonance $3p_{1/2}$ at  $A =124$, for instance,
has a width $\Gamma=61$ keV which is comparable to the
resonance energy $e_{\rm res} = 79$ keV, i.e., $\Gamma/e_{\rm res} \sim 1$.
This ratio decreases with $A$.  We find that the large deviation
is seen in the case of the large width-energy ratio
$\Gamma/e_{\rm res} \gesim 1/3$
(at $124 \le A \lesim 134$). The same is seen also for the $3p_{3/2}$
resonance. On the other hand, the small deviation can be linked to the
small width-energy ratio $\Gamma/e_{\rm res} \lesim 1/5$ as seen in the
$3p_{1/2}$ and $3p_{3/2}$ resonances at $A=136$ and $138$, and also in
the $2f_{5/2}$ and $2f_{7/2}$ resonances in all the isotopes.
Clearly the large width-energy ratio $\Gamma/e_{\rm res} \gesim 1/3$
indicates that the coupling to the continuum scattering states is
strong even though the resonance energy  is located below
the potential barrier ($e_{\rm res} < V_{\rm max}$ or $E <V_{B}$).
 Consequently the bound
state approximation does not work. For the same reason  it is
hard to describe these resonances in a box-discretized
calculation unless the box size is taken sufficiently
large.  In order to describe the resonance
at such a small energy $e_{\rm res} \sim 50$ keV, one has to use
a large box satisfying $\hbar^2/2m\times(\pi/R_{\text{box}})^2 \lesim
e_{\rm res}$, i.e., $R_{\rm box} \gesim 60$ fm.
A further larger box is
needed in order to describe the distribution around the peak~\cite{Bennaceur00} to evaluate the width, or one needs to combine
with other methods, such as stabilization method~\cite{Pei2011}, etc.

 From the above analysis, we can conclude that
 the correct treatment of the continuum is important
 to describe the quasiparticle resonances originating from
 weakly bound orbits with low angular momenta.

\section{\label{subsec:dendecay}Exponential tails of the pair correlated halo}

 In this subsection, we will discuss the
 asymptotic behavior of the density and the pair density
 associated with the giant neutron halo. We have already
 discussed in section III that the density and the pair density
 exhibit extended tails which can be characterized by
 an exponentially decreasing behavior as a function of $r$.
 In order to understand the origins of
the tail behaviors, we shall investigate the neutron density $\rho_{lj}(r)$
and the neutron pair density $\tilde{\rho}_{lj}(r)$ which are decomposed with
respect to the partial waves $lj$.

 {Figure 8 (a) and (b) shows the $lj$-decomposed neutron
 density $4\pi r^2 \rho_{lj}(r)$
 weighted with the volume element $4\pi r^2$ for $^{126}$Zr.}
We already saw the dominance of the $p$-wave component in the halo tail region ($r \gesim 10$ fm) in
connection with Fig.~\ref{fig:Zr-density-ratio} in
section III. Noticeably
the exponential slopes in the tail region are different for different
partial waves.
 The $p_{1/2}$ and $p_{3/2}$ components have the most gentle slopes
(hence they dominates in the tail) while the slopes of $f_{7/2}$ and  $f_{5/2}$ are
steeper than the $p$-waves. The densities arising
from the $s$- and $d$-waves are much steeper, and thus their
contributions to the halo tail is negligible.

{Looking at the $lj$-decomposed neutron pair densities $4\pi r^2\tilde{\rho}_{lj}(r)$,
shown in panels (c)  and (d),} we immediately see that the exponential slopes
of the neutron pair densities are apparently different from the slopes associated with
the neutron density $\rho_{lj}(r)$ (panels (a) and (b)),
and it is also obvious that the exponential slopes exhibit much weaker dependence on $lj$. The slopes
are rather similar among the partial waves $s_{1/2}, d_{3/2}$ and $d_{5/2}$ as well as
$p_{1/2},p_{3/2}, f_{5/2}$ and $f_{7/2}$.

\subsection{neutron density}

In order to clarify the behaviors of the exponential
slopes in the halo tail region, we shall quantify the exponential slope of
the neutron density $\rho_{lj}(r)$.
For this purpose, we consider a simple fitting function
\beq
{r^2}{\rho_{{\rm fit}, lj}(r) \propto  \left(r h_{l}^{(+)} (i\kappa_{lj}~r) \right)^2}
\label{eq:rhojl-fit}
\eeq
with $ h_{l}^{(+)}(z)$ being the spherical Hankel function \cite{Messiah}.
A concrete form for the orbital angular momentum $l=1$ is
 $r^2 \rho_{{\rm fit}, lj}(r)= {C} e^{-2\kappa_{lj}r}\left[ 1 + \frac{1}{\kappa_{lj}r}\right]^2$.
This is an asymptotic form expected to be obtained if the tail density is
contributed by a single quasiparticle state with a fixed energy.
{(Note that the asymptotic
form of the second-component wave function is
$\varphi_{2,lj}(r)/r \sim  h_{l}^{(+)}(i\kappa_{lj} r)$ with
$\kappa_{lj}=\sqrt{2m(E+|\lam|)/\hbar^2}$.)}
We fit this function $r^2\rho_{{\rm fit}, lj}(r)$ to
the numerically obtained density $r^2\rho_{lj}(r)$ in the tail
region, and we extract the parameter
$2\kappa_{lj}$ that represents the exponential slope of $r^2\rho_{lj}(r)$.
A simpler choice of the fitting function would be
$r^2\rho_{{\rm fit}, lj}(r)=C\exp(-2\kappa_{lj} r)$
valid for $\kappa_{lj} r \gg 1$.
But we do not adopt this because
the effect of the centrifugal barrier is not completely negligible
(i.e. $\kappa_{lj} r \gg 1$ is not realized) in  the region under discussion, $r \sim 20$ fm.
  We use the interval $r=15-20$ fm for the fit,
and we denote the extracted value as $2\kappa_{lj}$,
which we shall call as {\it
asymptotic exponential constant}.

{The extracted asymptotic exponential constant $2\kappa_{lj}$ is shown in
Fig.~\ref{fig:Zr-decayconst} (a) and (b)}. It is noticed that
the extracted values $2\kappa_{lj}$ differ for different $lj$, and their
isotopic dependencies are also different. For instance,
 $2\kappa_{lj}$ for $f_{5/2}$ is obviously different from that for
$f_{7/2}$ although
both equally have sizable contributions to the halo density.

We shall now show that the asymptotic exponential constants  $2\kappa_{lj}$
of these partial waves are governed by the low-lying
quasiparticle resonances $3p_{1/2}, 3p_{3/2}, 2f_{5/2}$ and $2f_{7/2}$.
To show this,  we evaluate
a part of the neutron density $\rho_{lj}(r)$, denoted by $\rho'_{lj}(r)$ below,  in which we consider only contributions
from the low-lying quasiparticles with $E=0-3$ MeV including the resonances mentioned
above.  The truncated neutron density $\rho'_{lj}(r)$ can be calculated by using Eq.~(\ref{eq:rhojl-GF})  in which the contour of the
integral encloses only the quasiparticle states with $E=0-3$ MeV. {In Fig.~\ref{fig:Zr138-rhojlrhojlint-e} (a)
we compare $\rho'_{lj}(r)$  with $\rho_{lj}(r)$ for $p_{3/2}$ contributions
in $^{138}$Zr.} It is seen that
$\rho'_{p_{3/2}}(r)$ reproduces $\rho_{p_{3/2}}(r)$ very nicely as far as the tail region
$r>10$~fm is concerned. We
extract the asymptotic exponential constant
$2\kap'_{lj}$ for $\rho'_{p_{3/2}}(r)$
by using the same
fitting function of Eq.~(\ref{eq:rhojl-fit}), and we confirm that
 $2\kap'_{lj}$ and $2\kap_{lj}$ are almost the same.
We also compare the tail of the neutron
density $\rho_{lj}(r)$ with that of the
density $\rho_{lj}(r,E_{\rm res})$ which represents
a contribution from the quasiparticle state
at the resonance energy $E_{\rm res}$ (cf. Table~\ref{tab:Egam-all}).
It is seen in Fig.~\ref{fig:Zr138-rhojlrhojlint-e}
 that the exponential slope of $\rho_{lj}(r,E_{\rm res})$
 agrees quite well with that of $\rho_{lj}(r)$.

To be more quantitative, we compare  in Fig.~\ref{fig:Zr-decayconst}~(b) the
extracted asymptotic exponential constant $2\kappa_{lj}$ of
$\rho_{lj}(r)$ with the exponential constant
\beq
{2\kappa_{lj,{\rm res}}}
  {= 2\sqrt{\frac{2m (E_{\rm res} + |\lam|)}{\hbar^2}}}
 \label{eq:2kap-res}
\eeq
{of $\rho_{lj}(r,E_{\rm res})$,} {which can be calculated  using
the resonance energy $E_{\rm res}$},
for the partial waves  $p_{1/2},p_{3/2}, f_{5/2}$ and
$f_{7/2}$ in all the isotopes.  The agreement is very good and
 it is even hard to see the difference between the two in some cases.
We thus confirm that the low-lying quasiparticle resonances
$3p_{1/2}, 3p_{3/2}, 2f_{5/2}$ and $2f_{7/2}$ corresponding to the
weakly-bound HF single-particle orbits dominate the tail and govern its
asymptotic exponential constant.

 The asymptotic exponential constants  $2\kappa_{lj}$ of the $s_{1/2},d_{3/2}$ and $d_{5/2}$
 partial wave neutron densities behave in a  different way as seen
in Fig.~\ref{fig:Zr-decayconst}~(a). Firstly, we observe
the large values of $2\kappa_{lj}$, which can be attributed to the fact
that the tail of $\rho_{lj}(r)$ in these partial waves is dominated by
deeply bound orbits.  Secondly we find that
the values of $2\kappa_{lj}$
decrease steeply with increasing $A$. This feature can
be explained by the fact that
the non-resonant continuum states in the low-lying
region of these partial waves
contribute more as the
neutron drip-line is approached. In
Fig.~\ref{fig:Zr138-rhojlrhojlint-e} (b) we plot the
truncated neutron density $\rho'_{lj}(r)$
of $s_{1/2}$ partial wave in $^{138}$Zr,
which represents the contributions from low-lying
non-resonant continuum states.
Although $\rho'_{lj}(r)$ is smaller than $\rho_{lj}(r)$ by
several orders,
it dominates the far external part ($r\gtrsim 15$~fm)
of the total density $\rho_{lj}(r)$.
Note, however,
that the $s_{1/2},d_{3/2}$ and $d_{5/2}$
densities do not influence the tail of the total neutron density as their
relative contributions are negligible in comparison with
those contributed from the low-lying $p,f$-wave
quasiparticle resonances.  Therefore, the complex behavior of $2\kap_{lj}$ in the $s,~d$-wave does not affect the conclusion on
the dominance of the low-lying quasiparticle
resonances in the exponential tail
of the neutron densities.

\subsection{neutron pair density}

Let us now examine the asymptotic exponential constants
of the
neutron pair densities $\tilde{\rho}_{lj}(r)$.
Similarly to the neutron density $\rho_{lj}(r)$, we
fit an exponentially decreasing function to the numerically obtained
neutron pair density $\tilde{\rho}_{lj}(r)$, but
we need a proper fitting function other than
$\rho_{{\rm fit},lj}(r)$ (Eq.(\ref{eq:rhojl-fit}))
which is appropriate only for
the normal density. We adopt
\beq
r^2\tilde{\rho}_{{\rm fit}, lj}(r) \propto  rh_{l}^{(+) }(i\tilde{\kappa}_{lj} r)
\eeq
as a fitting function.
Specifically, for $l=1$, we have
$r^2\tilde{\rho}_{{\rm fit}, lj}(r) =\tilde{C}  e^{-\tilde{\kappa}_{lj} r}\left[ 1 + \frac{1}{\tilde{\kappa}_{lj} r}\right]$.
This functional form is based on the following
considerations. We first point out that
there is no simple estimate for the asymptotic form of the pair density.
Even if we assume that
a quasiparticle state with energy $E$ contributes to the pair density,
the contribution
$ r^2\tilde{\rho}_{lj}(r,E) \propto \varphi_{1,lj}(r)\varphi_{2,lj}(r)$
does not have an exponential form since the first-component
wave function $\varphi_{1,lj}(r)$ is an oscillating sinusoidal function in the asymptotic region as the quasiparticle states under consideration
are all embedded above the threshold energy.  We therefore
exploit only the asymptotic form of $\varphi_{2,lj}(r)$, which behaves as $\varphi_{2,lj}(r) \sim rh_l^{(+)}(i\tilde{\kappa}_{lj}r)$,
 to prepare the fitting function. We use the same interval $r=15-20$ fm for the fitting, and denote the extracted asymptotic exponential constant as  $\tilde{\kappa}_{lj}$ for the pair density.

The extracted value of the asymptotic exponential constant $\tilde{\kappa}_{lj}$ is plotted in Fig.~\ref{fig:Zr-decayconst}~(c) and (d).
It is seen here that the behaviors of $\tilde{\kappa}_{lj}$
are different from those of the asymptotic exponential constants
$2 {\kappa}_{lj}$ of the neutron densities $\rho_{lj}(r)$. This immediately
demonstrates that
the quasiparticle resonances are not the major origin of the exponential tail
of the pair densities $\tilde{\rho}_{lj}(r)$. It is also seen that the values of
$\tilde{\kappa}_{lj}$ for all the partial waves $s_{1/2},p_{1/2},p_{3/2},
d_{3/2}, d_{5/2}, f_{5/2}$ and $f_{7/2}$
are rather similar, and they all  exhibit
similar isotopic
dependencies.

It is useful here to refer to the argument
in Ref.~\cite{DobHFB2} which asserts that
the asymptotic
exponential constant of the pair density may be given by
\beq
\tilde{\kappa}_{\rm min} = \sqrt{\frac{4m |\lam|} {\hbar^2}}\ \ .
\eeq
This is based on the consideration that
the asymptotic
behavior of the pair density is governed by the contribution of
the lowest energy quasiparticle state,
and in the case of nuclei near the
drip-line, the lowest energy quasiparticle state is the
continuum state with $E=|\lam|$ (cf. Eq.~(\ref{eq:2kap-res})).
In fact we see in
Fig.~\ref{fig:Zr-decayconst}~(c) and (d) that the isotopic
dependencies of $\tilde{\kappa}_{lj}$ has some resemblance to that of
 $\tilde{\kappa}_{\rm min}$ although a difference by a factor of about 2 is seen here. We thus deduce that non-resonant
 continuum quasiparticle states dominates
 the asymptotic tail of the pair density,
 although the difference by a factor of $\sim 2$
 suggests something beyond the argument in Ref.~\cite{DobHFB2}.

Figure~\ref{fig:Zr138-rhojlrhojlint-e} (c) and (d)
are made to check the above deduction.
In these panels we compare the
neutron pair density $\tilde{\rho}_{lj}(r)$ with
the truncated pair density
 $\tilde{\rho}'_{lj}(r)$ where only the low-lying quasiparticle states with
 $E<3$ MeV are included in the sum of Eq.~(\ref{eq:rhojl-GF}).
 It is also compared in panel (c) with
the pair density
  $\tilde{\rho}_{lj}(r,E_{\rm res})$ that originates from the single quasiparticle
  state at the resonance energy $E_{\rm res}$
  for $p_{3/2}$ partial wave. It is seen that
  $\tilde{\rho}_{lj}(r,E_{\rm res})$ apparently fails to describe the exponential tail
  since $\varphi_{1,lj}(r,E_{\rm res})$ and hence
$\tilde{\rho}_{lj}(r,E_{\rm res})$
oscillate with $r$. On the other hand the exponential tail of the pair density
 $\tilde{\rho}_{lj}(r)$ is nicely reproduced by the pair density
 $\tilde{\rho}'_{lj}(r)$ truncated with $E<3$ MeV.
This observation applies to the pair densities of all the $p,f$ and
 $s,d$ waves (see Fig.~\ref{fig:Zr138-rhojlrhojlint-e} (d) for $s_{1/2}$).
Recall that the $s,d$-waves have only non-resonant quasiparticle
 states in the low-lying spectrum $E<3$ MeV.
 We thus conclude that the non-resonant continuum quasiparticle states dominate the
 asymptotic exponential tail of the neutron pair density. It is in contrast to
 that of the neutron density in which the low-lying quasiparticle resonances dominate.
 It is also concluded that the exponential tail of the pair density is governed by
 coherent superpositions of the contributions from the non-resonant continuum quasiparticle
 states near the threshold energy. Under this situation,
 we can evaluate approximately the exponential tail
 arising from the non-resonant continuum
quasiparticle states as
\beq
  {r^2}\tilde{\rho}_{lj}(r) = \int_{|\lam|} dE \varphi_{1,lj}(r,E)\varphi_{2,lj}(r,E)
  \sim \int_0 de \sin(kr) e^{-\kappa r}
\eeq
with $k=\sqrt{2me/\hbar^2}$,  $\kappa=\sqrt{2m(2|\lam|+e)/\hbar^2}$ and  $e=E-|\lam|$.
{Since the weight factor $e^{-\kappa r}$ is dominant in an interval $0< e \lesim |\lam|$,
 the superposition of non-resonant continuum wave
 $\sin(kr)$ with such a weight factor gives $\sim e^{-\tilde{\kappa}_{{\rm nr}}r}$
with $\tilde{\kappa}_{{\rm nr}} \sim O(\tilde{\kappa}_{{\rm min}})$.  Consequently we can expect
$\tilde{\rho}_{lj}(r) \sim e^{-\tilde{\kappa}_{lj} r}$ with $\tilde{\kappa}_{lj} =\tilde{\kappa}_{{\rm min}}+\tilde{\kappa}_{{\rm nr}}
\sim 2 \tilde{\kappa}_{{\rm min}}$.} This estimation is in qualitative agreement with the
observation $\tilde{\kappa}_{lj} \approx 2 \tilde{\kappa}_{{\rm min}}$ seen in the extracted
asymptotic exponential constant $\tilde{\kappa}_{lj}$ (cf. Fig.~\ref{fig:Zr-decayconst}~(c) and (d)).



\section{\label{summary}Conclusions}


We have investigated the pair correlation in
 $^{124-138}$Zr by using the self-consistent
continuum Skyrme Hartree-Fock-Bogoliubov (HFB) approach,
in which the asymptotic behaviors of the wave functions of the
continuum quasiparticle states are properly treated with the Green's function method.
The giant neutron halos in
these very neutron-rich weakly-bound nuclei
are analyzed in detail.
Focuses are put on properties of the halo part of the neutron pair condensate,
i.e. the exponential tail of
the neutron pair density $\tilde{\rho}(r)$, in particular, the r.m.s. radius,
its single-particle composition, and the exponential slope parameter
(the asymptotic exponential constant).  We found that
these  are apparently different from the corresponding quantities
associated with
 the exponential tail of the neutron density ${\rho}(r)$.

In order to
clarify the origin of the differences, we have looked into
the spectrum of the neutron quasiparticle states. 
Because of the small Fermi energy $\lam$
and of the relative largeness of the pairing gap $\Delta \gesim |\lam|$,
all the quasiparticle states in $^{126-138}$Zr are located above the threshold
energy. Typical examples are
the quasiparticle resonances
$3p_{1/2}, 3p_{3/2}$ and $2f_{7/2}$, which play
central roles
in forming the giant neutron halo.
 They  originate from
the weakly bound orbits in the Hartree-Fock (HF) potential,
but they appear as
resonances with finite width once the
pair correlation is taken into account.
There exist non-resonant continuum quasiparticle states, and they
contribute also to the giant halo. Such quasiparticle
states are seen in all the partial waves, and typically in
$s_{1/2},d_{3/2}$ and $d_{5/2}$.

The central finding in the present analysis is that
the exponential  tail of the neutron pair density $\tilde{\rho}_{lj}(r)$
is contributed mainly from  the low-lying non-resonant continuum
quasiparticle states.
This is contrasting to the microscopic structure of
the exponential tail of the neuron density  ${\rho}_{lj}(r)$, the main
contributions of which are from
the low-lying quasiparticle resonances $3p_{1/2}, 3p_{3/2}$ and $2f_{7/2}$
as well as  $2f_{5/2}$, all
corresponding to the {HF} orbits and resonances near the Fermi energy.

The different microscopic origins
of the neutron density  ${\rho}_{lj}(r)$ and of the neutron pair
density $\tilde{\rho}_{lj}(r)$
reflect to the asymptotic exponential
constants  $2\kappa_{lj}$ and $\tilde{\kappa}_{lj}$
parameterizing the exponential slope of the tail
of the neutron density  ${\rho}_{lj}(r)$ and  the
neutron pair density  $\tilde{\rho}_{lj}(r)$, respectively.
Concerning the  neutron density  ${\rho}_{lj}(r)$ of the
dominant partial waves
$p_{1/2}$ and $p_{3/2}$, subdominant $f_{5/2}$ and $f_{7/2}$, we found that
the constant $2\kappa_{lj}$ is governed
essentially  by the peak energy of the lowest-lying quasiparticle resonance
that arises from the bound or resonant HF single-particle orbit closest to
the Fermi energy.  The asymptotic exponential constant $\tilde{\kappa}_{lj}$
of the neutron pair density $\tilde{\rho}_{lj}(r)$ has
a close relation to the threshold energy $|\lam|$ of
the non-resonant continuum states, rather than
 the energies of the low-lying quasiparticle resonances,
 since the superpositions of the non-resonant continuum
 quasiparticle states dominate the tail of $\tilde{\rho}_{lj}(r)$.

In the present analysis, we have also put a focus on the comparison between
 the continuum HFB calculation with the box-discretized
HFB calculation. This comparison has pointed to a few specific problems of
the box calculation,
which arise if the size of the box is not sufficiently large. One
is that
 the finite box causes relatively large error in describing
 the exponential halo tail of
the neutron pair density  $\tilde{\rho}_{lj}(r)$, in comparison with the
tail of the neutron density ${\rho}_{lj}(r)$. This is because the halo tail
of the neutron pair density  $\tilde{\rho}_{lj}(r)$, which is dominated by
the contributions of the non-resonant continuum quasiparticle states with
small asymptotic kinetic energies $e=E-|\lam|$, is more extended than the
neutron density ${\rho}_{lj}(r)$.
Another problem is found in
describing the quasiparticle resonances corresponding to the
weakly-bound $3p_{1/2}$ and $3p_{3/2}$ orbits, in particular in
the isotopes with $A \lesim 132$. We here have only
small centrifugal barriers in the $p$-waves, which
makes the coupling to the scattering states strong. Consequently,
the quasiparticle resonances $3p_{1/2}$ and $3p_{3/2}$ are not
represented well by any eigen states in
the discretized calculation with a small box.

\begin{acknowledgments}

This work was partly supported by the Major State 973 Program 2013CB834400;
the National Natural Science Foundation of China under Grants No.
10975007, No. 10975008, No. 11005069, and No. 11175002;  the Research Fund for the
Doctoral Program of Higher Education under Grant No. 20110001110087;  the Oversea
Distinguished Professor Project from Ministry of Education No. MS2010BJDX001, and the
Grant-in-Aid for Scientific Research (Nos. 21340073 and 23540294) from the Japan Society
for the Promotion of Science.

\end{acknowledgments}


\begin{thebibliography}{100}
\bibitem{Tanihata} I.~Tanihata, \textit{et al.}, Phys. Rev. Lett. {\bf 55}, 2676 (1985).

\bibitem{Meng-Zr}
J.~Meng and P.~Ring, Phys. Rev. Lett. {\bf 80}, 460 (1998).

\bibitem{DobHFB1}
J.~Dobaczewski, H.~Flocard and J.~Treiner, Nucl. Phys. A {\bf 422}, 103 (1984).

\bibitem{DobHFB2}
J.~Dobaczewski, W.~Nazarewicz, T.~R.~Werner, J.~F.~Berger,
C.~R.~Chinn, and J.~Decharg\'e, Phys. Rev. C {\bf 53}, 2809 (1996).

{\bibitem{Bender2003}
M. Bender, P.-H. Heenen, and P.-G. Reinhard,
Rev. Mod. Phys. {\bf 75}, 121 (2003).}

\bibitem{Bulgac}
A.~Bulgac, preprint FT-194-1980, Bucharest, 1980, nucl-th/9907088.

\bibitem{Meng-Li} J.~Meng and P.~Ring, Phys. Rev. Lett. {\bf 77}, 3963
(1996).

{\bibitem{Vretenar2005} D. Vretenar, A. V. Afanasjev, G. A. Lalazissis, and P. Ring,
Phys. Rep. {\bf 409}, 101~(2005).}

\bibitem{Meng2006} J. Meng, H. Toki, S.-G. Zhou, S. Q. Zhang, W. H. Long, and L. S. Geng, Prog. Part. Nucl. Phys. {\bf 57}, 470~(2006).

\bibitem{MengRCHB1998}
J.~ Meng, Nucl. Phys. A {\bf 635}, 3 (1998).

\bibitem{ZhangSQ2003} S. Q. Zhang, J. Meng, and S.-G. Zhou, Sci. China Ser. G  {\bf 46}, 632~(2003).

\bibitem{Meng-Ca}
J. Meng, H. Toki, J. Y. Zeng, S. Q. Zhang, and S.-G. Zhou, Phys. Rev. C {\bf 65}, 041302(R) (2002).

\bibitem{Terasaki2006} J. Terasaki, S. Q. Zhang, S.-G. Zhou, and J. Meng, Phys. Rev. C {\bf 74}, 054318~(2006).

\bibitem{Zhou2003} S.-G. Zhou, J. Meng, and P. Ring, Phys. Rev. C {\bf 68}, 034323~(2003).

\bibitem{Zhou2010} S.-G Zhou, J. Meng, P. Ring, and E.-G. Zhao, Phys. Rev. C {\bf 82}, 011301(R) ~(2010).

\bibitem{Long2010} W. H. Long, P. Ring, N. Van Giai, and J. Meng, Phys. Rev. C {\bf 81}, 024308~(2010).

\bibitem{LiLuLu2012} L.~L.~Li, J. Meng, P. Ring, E.-G. Zhao,  and S.-G Zhou, Phys. Rev. C {\bf 85}, 024312
(2012).

\bibitem{ChenY2012} Y.~Chen, L.~L.~Li, H.~Z.~Liang, and J.~Meng, Phys. Rev. C {\bf 85}, 067301
(2012).

\bibitem{Bennaceur00}
K.~Bennaceur, J.~Dobaczewski, and M.~Ploszajczak,
Phys. Lett. B {\bf 496}, 154~(2000).

\bibitem{DD-Dob}
J.~Dobaczewski, W.~Nazarewicz, and P.-G.~Reinhard,
Nucl. Phys. A {\bf 693}, 361 (2001).

\bibitem{Grasso}
M.~Grasso, N.~Sandulescu, N. Van Giai and R.~J.~Liotta,
Phys. Rev. C {\bf 64}, 064321 (2001).

\bibitem{Hamamoto03}
I. Hamamoto and B. R. Mottelson, Phys. Rev. C {\bf 68}, 034312~(2003).

\bibitem{Hamamoto04}
I. Hamamoto and B. R. Mottelson, Phys. Rev. C {\bf 69}, 064302~(2004).

\bibitem{MMS05}
M.~Matsuo, K.~Mizuyama, and Y.~Serizawa,
Phys. Rev. C {\bf 71},  064326 (2005).

\bibitem{Yamagami05}
M.~Yamagami, Phys. Rev. C {\bf 72}, 064308 (2005).  

\bibitem{GrassoGhalo}
M.~Grasso, S.~Yoshida, N.~Sandulescu, and N. Van Giai,
Phys. Rev. C {\bf 74}, 064317 (2006).


\bibitem{Oba}
H.~Oba and M.~Matsuo, Phys. Rev. C {\bf 80}, 024301 (2009).

\bibitem{ZhangY2011} Y. Zhang, M. Matsuo, and J. Meng,
Phys. Rev. C {\bf 83}, 054301 (2011).

\bibitem{Hagino11}
K.~Hagino and H.~Sagawa, Phys. Rev. C {\bf 84}, 011303(R) (2011).


\bibitem{Long2010Ce} W. H. Long, P. Ring, J. Meng, N. Van Giai, and C. A. Bertulani, Phys. Rev. C {\bf 81}, 031302(R)~(2010).

\bibitem{Belyaev}
S.~T.~Belyaev, A.~V.~Smirnov, S.~V.~Tolokonnikov and S.~A.~Fayans,
Sov. J. Nucl. Phys. {\bf 45}, 783 (1987).



\bibitem{YangSC} S.~C.~Yang, J.~Meng and S.-G.~Zhou, Chinese Phys. Lett {\bf 18},196~(2001).

\bibitem{ZhangSS} S.~S.~Zhang, J.~Meng, S.-G.~Zhou, and
G.~C.~Hillhouse, Phys. Rev. C {\bf 70}, 034308 (2004).

\bibitem{Zhang2008} L. Zhang, S.-G. Zhou, J. Meng, and E. G. Zhao, Phys. Rev. C {\bf 77}, 014312~(2008).

\bibitem{Pei2011} J. C. Pei, A. T. Kruppa, and W. Nazarewicz, Phys. Rev. C {\bf 84}, 024311~(2011).


\bibitem{Michel2008} N. Michel, K. Matsuyanagi, and M. Stoitsov,  Phys. Rev. C {\bf 78}, 044319~(2008).

\bibitem{Messiah}A. Messiah, {\em Quantum Mechanics},
North Holland, Amsterdam, 1962.

\bibitem{SkI4} E. Chabanat, P. Bonche, P. Haensel, J. Meyer, and R. Schaeffer, Nucl. Phys. A {\bf 635}, 231~(1998).

\bibitem{Meng1998PRC} J. Meng, Phys. Rev. C {\bf 57}, 1229 (1998).

\bibitem{Matsuo2006PRC} M. Matsuo, Phys. Rev. C {\bf 73}, 044309~(2006).

\bibitem{Matsuo2007NPAc} M. Matsuo, Y. Serizawa, and K. Mizuyama, Nucl. Phys. A {\bf 788}, 307c~(2007).

\bibitem{Matsuo2010PRC} M. Matsuo and Y. Serizawa, Phys. Rev. C {\bf 82}, 024318~(2010).

{
\bibitem{Shimoyama2011} H. Shimoyama and M. Matsuo, Phys. Rev. C {\bf 84}, 044317~(2011).}

\end{thebibliography}

\clearpage

%
%
%

%

%

%
%

\begin{figure}[htb]
\includegraphics[width=15cm]{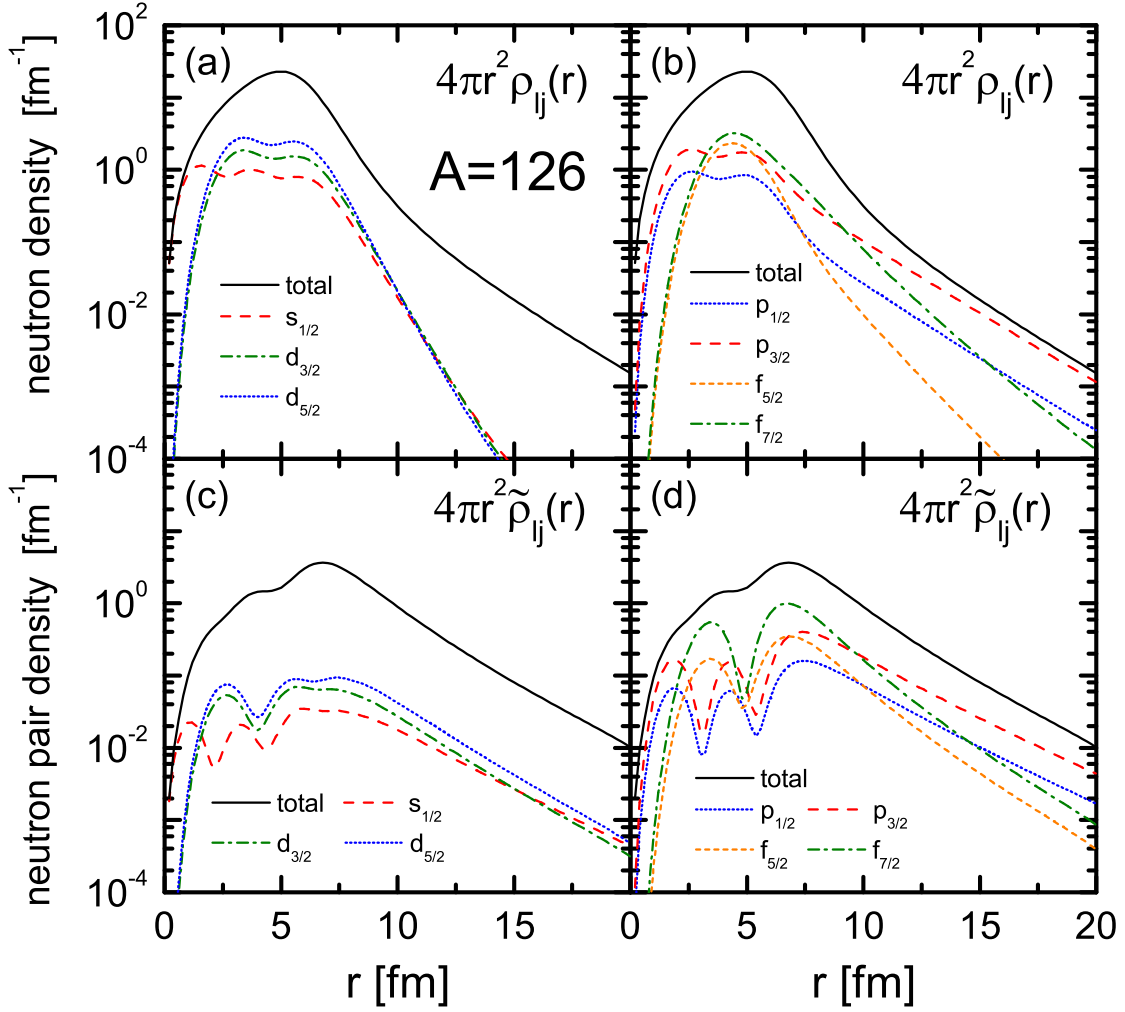}
\caption{(a)-(b) Neutron density $4\pi r^2{\rho}_{lj}(r)$ and (c)-(d)
neutron pair density $4\pi r^2\tilde{\rho}_{lj}(r)$ of the $s_{1/2},~p_{1/2},~p_{3/2},~d_{3/2},~d_{5/2},~f_{5/2}$, and $f_{7/2}$
partial waves in $^{126}$Zr.
The total neutron density $4\pi r^2{\rho}(r)$
and neutron pair density $4\pi r^2\tilde{\rho}(r)$
are also plotted with the solid line.
  }\label{fig:Zr126-rhorhotjl}
\end{figure}

\begin{figure}[htb]
\includegraphics[width=15cm]{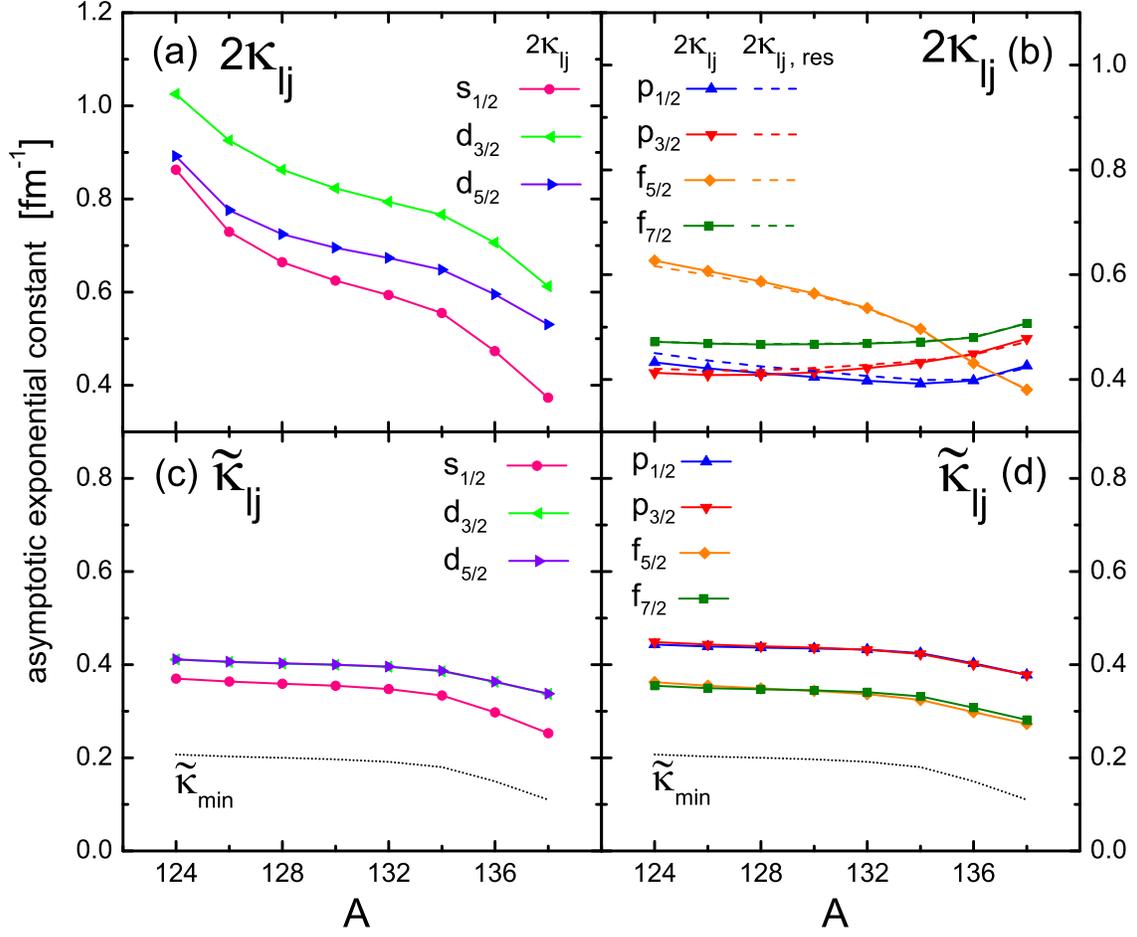}
\caption{(a) (b) Asymptotic exponential constant $2\kap_{lj}$ of the neutron density $\rho_{lj}(r)$,
 and (c) (d) $\tilde{\kap}_{lj}$ of the neutron pair density $\tilde{\rho}_{lj}(r)$
 for the $s_{1/2},~p_{1/2},~p_{3/2},~d_{3/2},~d_{5/2},~f_{5/2}$, and $f_{7/2}$ partial waves in the Zr isotopes.
 The fitting interval is $r=15\sim 20$~fm.
 The estimated asymptotic exponential constant
 $2\kap_{lj,\text{res}}$ and $\tilde{\kap}_{\text{min}}$ are
 also shown with the dashed line and
 {dotted } line, respectively,
 where $\tilde{\kap}_{\text{min}}=\sqrt{4m|\lam|}/\hbar$ and
 ${\kap}_{lj,\text{res}}=\sqrt{2m(E_{\rm res}+|\lam|)/\hbar^2}$
 are evaluated using the Fermi energy $\lam$ and
 the resonance energy $E_{\rm res}$ shown in Table~\ref{tab:Egam-all}.
}\label{fig:Zr-decayconst}
\end{figure}

\begin{figure}[htb]
\includegraphics[width=15cm]{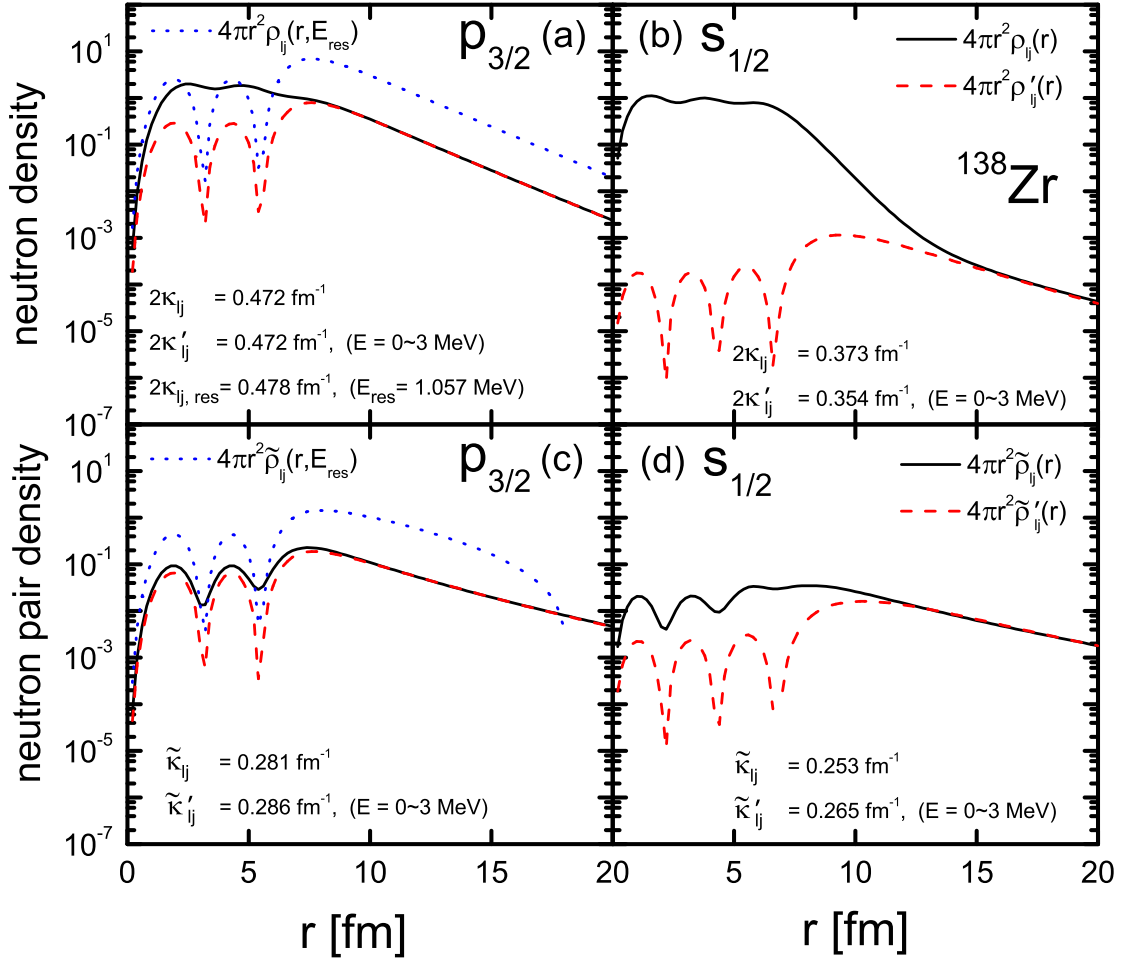}
\caption{(a) (b) Neutron density $4\pi r^2\rho_{lj}(r)$ (solid line),
and the truncated neutron density $4\pi r^2\rho'_{lj}(r)$
contributed from the low lying quasiparticle
states with $E<3$~MeV (dashed line)
(a) for $p_{3/2}$ partial wave
and (b) for $s_{1/2}$ partial wave in $^{138}$Zr.
The contribution $4\pi r^2\rho_{lj}(r,E_{\rm res})$
of the quasiparticle state at the resonance energy
$E_{\rm res}=1.057$~MeV is also shown (dotted line) for $p_{3/2}$
in panel (a).
The asymptotic exponential constant $2\kap_{lj}$ fitted to
the neutron density $4\pi r^2\rho_{lj}(r)$, $2\kap'_{lj}$
fitted to $4\pi r^2\rho'_{lj}(r)$, and $2\kap_{lj,{\rm res}}$
calculated with Eq.~(\ref{eq:2kap-res}) for $p_{3/2}$ partial wave are
labeled in the corresponding panels.
(c) (d) The same as (a) (b) but for the neutron pair densities
$4\pi r^2\tilde{\rho}_{lj}(r)$, $4\pi r^2\tilde{\rho}'_{lj}(r)$
and $4\pi r^2\tilde{\rho}_{lj}(r,E_{\rm res})$.
Note that the unit of the vertical axis is $[{\rm fm}^{-1}]$
for $4\pi r^2\rho_{lj}(r)$, $4\pi r^2\rho'_{lj}(r)$,
$4\pi r^2\tilde{\rho}_{lj}(r)$ and $4\pi r^2\tilde{\rho}'_{lj}(r)$,
but $[{\rm fm}^{-1}~{\rm MeV}^{-1}]$ for
$4\pi r^2\rho_{lj}(r,E_{\rm res})$ and $4\pi r^2\tilde{\rho}_{lj}(r,E_{\rm res})$.
  }
\label{fig:Zr138-rhojlrhojlint-e}
\end{figure}


\clearpage
\begin{table}
  \centering
  \caption{Fermi energy $\lam$ and the average pairing gaps $\Del_{uv}$ and $\Del_{vv}$
  for the Zr isotopes.  Listed are also barrier height
  of the Hartree-Fock (HF) plus centrifugal potential $V_{\rm max}$,
  HF single-particle energies $\vep$,
  resonance energies $E_{\rm res}$ and
  $e_{\rm res}=E_{\rm res}- |\lam|= E_{\rm res}+ \lam$, and width $\Gamma$ of
  quasiparticle resonances
  around the Fermi energy for each isotope. All in MeV.}\label{tab:Egam-all}
\begin{tabular}{cccccccccc}
  \hline \hline
             & $A$           & $124$   & $126$    & $128$    & $130$    & $132$    & $134$    & $136$    & $138$ \\ \hline
             & $\lam$        & $-0.446$& $-0.427$ & $-0.415$ & $-0.401$ & $-0.380$ & $-0.336$ & $-0.232$ & $-0.126$ \\
             & $\Del_{uv}$   & $0.468$ & $0.596$  & $0.656$  & $0.678$  & $0.667$  & $0.628$  & $0.585$  & $0.619$ \\
             & $\Del_{vv}$   & $0.416$ & $0.532$  & $0.589$  & $0.611$  & $0.605$  & $0.574$  & $0.539$  & $0.577$ \\ \hline
  $3p_{1/2}$ & $V_{\rm max}$ & $0.289$ & $0.260$  & $0.237$  & $0.220$  & $0.206$  & $0.195$  & $0.187$  & $0.184$ \\
             & $\vep$        & $0.134$ & $0.012$  & $-0.114$ & $-0.244$ & $-0.377$ & $-0.513$ & $-0.656$ & $-0.810$ \\
             & $E_{\rm res}$           & $0.524$ & $0.491$  & $0.466$  & $0.447$  & $0.438$  & $0.461$  & $0.589$  & $0.815$ \\
             & $e_{\rm res}$      & $0.079$ & $0.064$  & $0.051$  & $0.045$  & $0.058$  & $0.125$  & $0.357$  & $0.690$ \\
             & $\Gamma$      & $0.061$ & $0.042$  & $0.028$  & $0.021$  & $0.024$  & $0.044$  & $0.072$  & $0.083$ \\ \hline
  $3p_{3/2}$ & $V_{\rm max}$ & $0.289$ & $0.260 $ & $0.237 $ & $0.220 $ & $0.206 $ & $0.195 $ & $0.187 $ & $0.184 $ \\
             & $\vep$        & $-0.074$& $-0.209$ & $-0.346$ & $-0.485$ & $-0.625$ & $-0.766$ & $-0.911$ & $-1.069$ \\
             & $E_{\rm res}$           & $0.436$ & $0.437 $ & $0.452 $ & $0.484 $ & $0.540 $ & $0.632 $ & $0.811 $ & $1.057 $ \\
             & $e_{\rm res}$      & $-0.010$& $0.009 $ & $0.037 $ & $0.083 $ & $0.159 $ & $0.296 $ & $0.579 $ & $0.931 $ \\
             & $\Gamma$      & $-$     & $0.002 $ & $0.013 $ & $0.033 $ & $0.054 $ & $0.068 $ & $0.071 $ & $0.075 $ \\ \hline
  $2f_{5/2}$ & $V_{\rm max}$ & $2.418$ & $2.304 $ & $2.198 $ & $2.100 $ & $2.011 $ & $1.931 $ & $1.860 $ & $1.810 $ \\
             & $\vep$        & $1.112$ & $0.973$  & $0.831$  & $0.687$  & $0.541$  & $0.393$  & $0.238$  & $0.058$ \\
             & $E_{\rm res}$           & $1.590$ & $1.482 $ & $1.370 $ & $1.248 $ & $1.109 $ & $0.941 $ & $0.732 $ & $0.624 $ \\
             & $e_{\rm res}$      & $1.144$ & $1.054 $ & $0.955 $ & $0.846 $ & $0.729 $ & $0.605 $ & $0.500 $ & $0.498 $ \\
             & $\Gamma$      & $0.160$ & $0.130 $ & $0.100 $ & $0.072 $ & $0.047 $ & $0.027 $ & $0.015 $ & $0.012 $ \\ \hline
  $2f_{7/2}$ & $V_{\rm max}$ & $2.416$ & $2.302 $ & $2.197 $ & $2.099 $ & $2.009 $ & $1.930 $ & $1.859 $ & $1.809 $ \\
             & $\vep$        & $0.066$ & $-0.106$ & $-0.279$ & $-0.452$ & $-0.624$ & $-0.796$ & $-0.968$ & $-1.148$ \\
             & $E_{\rm res}$           & $0.708$ & $0.709 $ & $0.715 $ & $0.730 $ & $0.758 $ & $0.816 $ & $0.963 $ & $1.207 $ \\
             & $e_{\rm res}$         & $0.262$ & $0.282 $ & $0.301 $ & $0.328 $ & $0.378 $ & $0.480 $ & $0.731 $ & $1.081 $ \\
             & $\Gamma$      & $0.001$ & $0.001 $ & $0.002 $ & $0.002 $ & $0.002 $ & $0.003 $ & $0.006 $ & $0.012 $ \\
  \hline
\end{tabular}
\end{table}

\end{document}